\documentclass[fleqn,usenatbib]{mnras}

\usepackage{newtxtext,newtxmath}

\usepackage[T1]{fontenc}
\usepackage{ae,aecompl}

\usepackage{graphicx}	
\usepackage{amsmath}	
\usepackage{amssymb}	

\title[Failed prominence eruptions] {Failed prominence eruptions near 24 cycle maximum }

\author[Boris Filippov]{
B. Filippov \thanks{E-mail:
bfilip@izmiran.ru} 
\\
 Pushkov Institute of Terrestrial Magnetism,
Ionosphere and Radio Wave Propagation of the Russian Academy of
Sciences (IZMIRAN), \\ Troitsk, Moscow 108840, Russia}

\date{Accepted XXX. Received YYY; in original form ZZZ}

\pubyear{2015}

\begin{document}
\label{firstpage}
\pagerange{\pageref{firstpage}--\pageref{lastpage}}
\maketitle

\begin{abstract}
We analyze 16 failed filament eruptions observed near 24 solar cycle maximum from May 2013 to July 2014. No significant rotation of filament spines is observed during the ascent in all studied failed eruptions, which does not support kink-instability mechanism of triggering the eruptions. We calculate potential magnetic field distributions in the corona above the initial locations of the filaments to study their height dependence. In seven events, the vertical profiles of the decay index $n$ are monotonic. The other nine events occur in the regions with the switchback or saddle-like $n$-profiles. The direction of the horizontal field near the saddle bottom is turned through more than 100$^\circ$ relative its direction at the initial filament position, which reveals the quadrupolar magnetic configuration with null points in these regions. The eruptive filaments stop above the null points where the total Lorenz force is directed upward. The most reasonable force that can terminate filament ascending and balance the Lorenz force seems the gravity.

\end{abstract}




\section{Introduction}

Solar prominences, or filaments as they are called when observed in projection against the solar disc, sometimes suddenly start to ascend as a whole (full eruptions) \citep{Jo11,Ho15}  or within limited sections of their length (partial eruptions) \citep{Gi06,Kl14}. An eruptive prominence can move high into the corona (successful eruptions) and initiate a coronal mass ejection (CME). It is not uncommon that the erupting prominence decelerates and stops at some greater height in the corona (confined or failed eruptions) \citep{Ji03,To05,Al06,Ku13,Ku15}. A CME is not observed after such a 'prominence jump'. Therefore, observations of the beginning of the prominence rapid ascent, or in other words its activation \citep{Gi00}, do not ensure strong disturbance of space weather associated with CMEs. In the catalog of prominence eruptions \citep{Mc15}  observed by the Atmospheric Imaging Assembly (AIA) \citep{Le12} onboard the {\it Solar Dynamics Observatory} {\it SDO}) \citep{Pe12}   in a period from June 2010 to September 2014 (more than 900 events), about 20\% of all eruptions are classified as confined ones. 

Instabilities of magnetic flux ropes are considered as one of the most probable driving mechanisms of eruptive prominences. The twisted structure is often observed in eruptive prominences and cores of CMEs in the field-of-view of spaceborn coronagraphs \citep{Ga04,Fi08,Pa13,Jo14,Ch14}. The flux-rope magnetic structure is found in space in magnetic clouds arriving to the Earth's orbit after launches of CMEs \citep{Le90,Da07}. If twist of the flux-rope field lines exceeds some critical value, the flux rope axis deforms into a helix as a result of the kink instability \citep{Ka66,To04}. If a flux rope is embedded into magnetic field with a rather strong gradient, a catastrophic loss of equilibrium or torus instability can develop \citep{Ba78,Fo91,Fo95,Kl06,Lo14}. Kink instability is able to change the shape and position of the flux rope but it remains confined. Thus the kink instability can be a mechanism of failed eruptions. On the other hand, kinking may convey the flux rope to the region where the catastrophic loss of equilibrium or torus instability acts.

The role of the strapping magnetic field in the fate of an activated filament is widely accepted. In many studies, the so named magnetic decay index $n$ \citep{Ba78,Fi00,Fi01,Kl06} 
  \begin{equation}
n = - \frac{\partial \ln B_e}{\partial \ln h},
\end{equation} 
where $B_e$ is the horizontal coronal magnetic field and $h$ is the height above the photosphere, is calculated and analyzed. \citet{Gu11} found that the driven by the kink instability flux rope, related to the eruptive filament on 2005 May 27, was confined in the corona because the decay index of the external magnetic field stays below the threshold for the torus instability ($n_c$ = 1.5) within a long height range. The decay index was also lower the torus instability threshold at the maximum filament height in the failed eruption on 2011 March 9 \citep{Li18}. \citet{Liu08} reported that the average decay index in the height range 42 -- 105 Mm over source regions of ten eruptive events was consistently smaller ($n \leq 1.71$) for failed eruption than those for full eruptions ($n > 1.74$). The confinement of eruptions of two small-scale filaments on 2014 April 7 is attributed to the large-scale overlaying magnetic loops \citep{Xu16}.  The decay index was about 1.2 on average above the filaments, which was less than the threshold for the torus instability. 

However, sometimes eruptive filaments terminate ascending motion in regions where the decay index is greater that the threshold for the torus instability. \citet{Zh19} found that in seven cases from studied 16 the decay index at the filament-apex position exceeds the threshold for the torus instability. All these seven events demonstrate strong writhing motions during the eruptions with the rotation angle in the range of 50$^\circ$ -- 130$^\circ$. The other filaments stopped in torus-stable regions but while five of them showed rotation less than 20$^\circ$, four filaments also rotated through an angle in the range of 50$^\circ$ -- 120$^\circ$. Nevertheless, the authors insist that the rotation motion of a filament has a certain correlation with the failed eruption.

Some other factors are considered as supporting the confinement of eruptions. \citet{Liu09} suggested that the asymmetry of background fields provides a relatively stronger confinement for flux-rope eruptions than the symmetric fields do. Observations of EUV arcades above an erupting filament provide evidence that the overlying magnetic arcades could confine the filament eruption \citep{Ch13}. The amount of stored free magnetic energy in the region before the eruption can be important for the success or failure of eruptions \citep{Liu18}. Strong ambient toroidal magnetic field (along the flux rope axis) can prevent the flux rope from kinking and produce a toroidal tension force that halts the eruption \citep{My15,My16,Fi20}. Coronal magnetic field-line reconnection can play an important role in the creation of conditions for filament eruptions either strengthening or weakening the magnetic confinement of the filaments \citep{Wa18}.

Magnetic-field structure above active regions is analyzed also to discriminate them regarding the probability of eruptive or confined flares. \citet{Ch11} studied the properties of the magnetic field for six confined and three eruptive flares in the same active region NOAA 10720 on 2005 January. The vertical decay index profiles were monotonic for confined flares and saddle-like with a 'bump' at a height of about 10 Mm for eruptive flares. Moreover, the field strength for the eruptive flares was weaker than that for the confined events at the same heights.\citet{Wa17} found that the decay index increased monotonously with height in the majority ($\sim$ 85\%) of studied 60 flaring regions and reached the critical value for the torus instability ($n_c$ = 1.5) at significantly higher heights for confined flares than for eruptive ones. Saddle-like $n$-profiles in the other regions had significantly smaller values of $n$ at the saddle bottom in confined flares than that in eruptive ones. In addition to decay index calculations, which were in general in agreement with the previous results, \citet{ba18} analyzed the orientation of the magnetic field as a function of height in the corona. In the confined-flare regions, the field rotated more rapidly than in eruptive ones and quickly adjusted to the large-scale active-region field. The decay index in the flare-rich but CME-poor active region NOAA 12192 (2014 October) was lower than in CME-productive regions; $n$ did not reach a value of 1.5 until a large height of more than 120\,Mm, so the torus instability was unlikely \citep{Su15}. The decisive role of the decay index height-dependence for the active-region CME-productivity was confirmed in many other flares\citep{Va18,Sa18}.

In this paper, we analyze 16 failed filament eruptions observed near 24 solar maximum. We calculate potential magnetic field distributions in the corona above the initial locations of the filaments. The values of the decay index at the initial height of filaments before the eruptions and at the maximum observed height were found. The environment magnetic conditions at the heights where the filaments were stopped cannot be assumed as favourable for a stable equilibrium of a magnetic flux rope. We analyze a simple model for the flux-rope equilibrium and show the importance of the gravity force for the confinement of the eruptions.

\begin{figure*}
		\includegraphics[width=180mm]{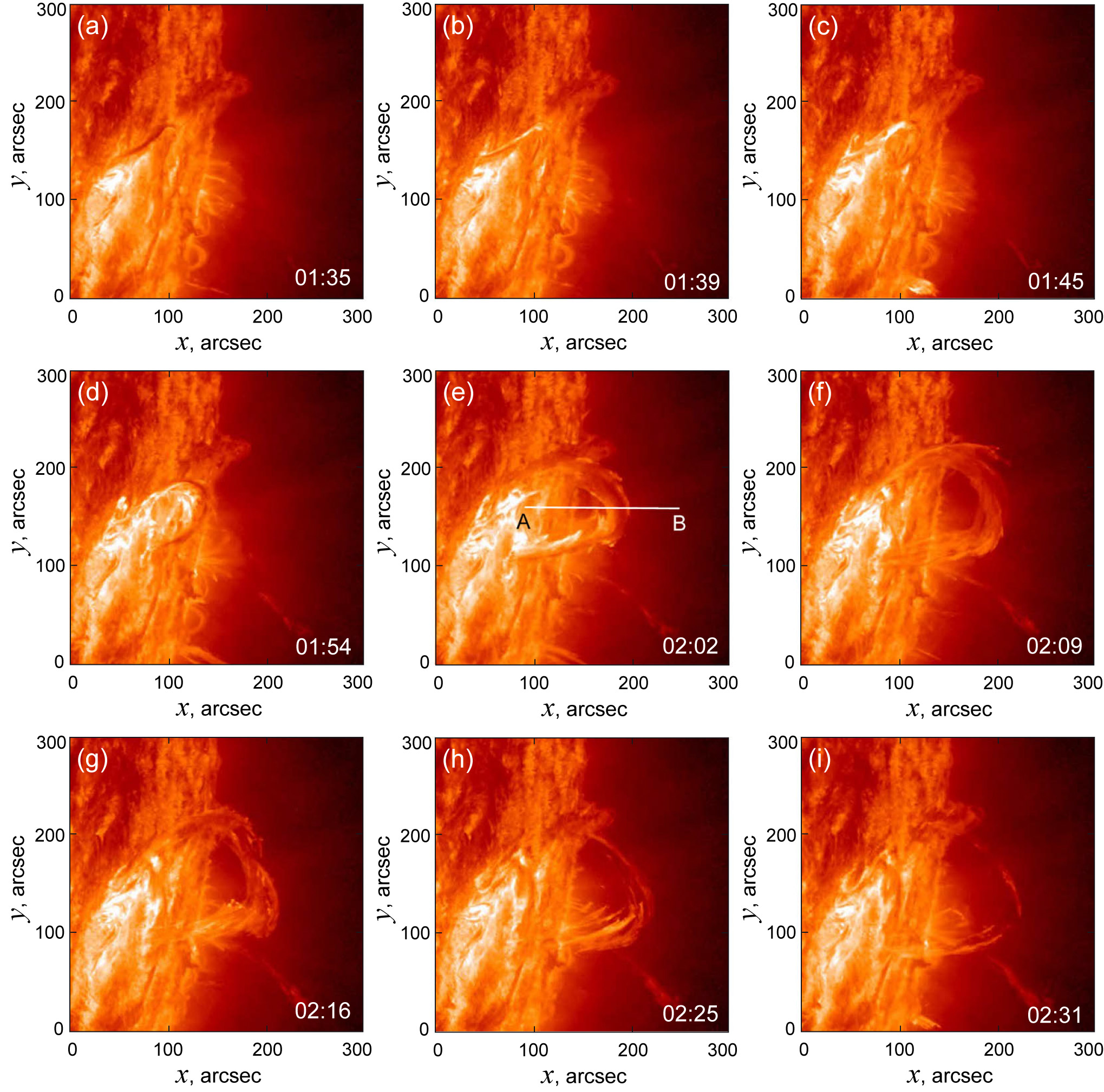}
    \caption{Failed filament eruption observed on 2014 March 28 by {\it SDO}/AIA in the 304-\AA\ channel close to the west limb. The white line AB in the panel (e) shows the slit position for the distance-time diagram presented in Fig. 2. (Courtesy of the NASA/{\it SDO} and the AIA science team.) }
    \label{fig:1}
\end{figure*}

\begin{figure}
		\includegraphics[width=90mm]{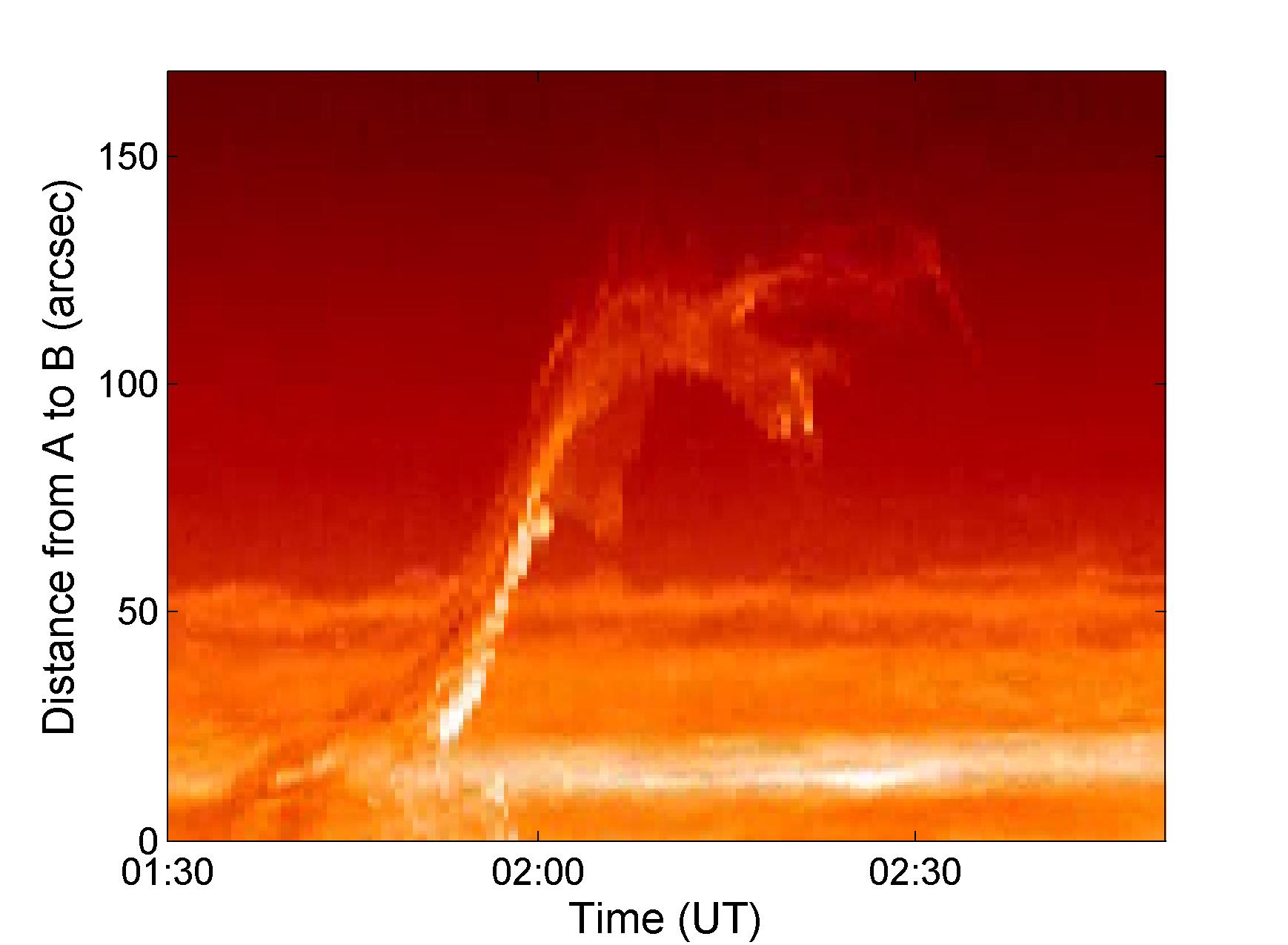}
    \caption{Distance-time diagram along the slit AB shown in Fig. 1e.}
    \label{fig:2}
\end{figure}

\section{Observational data}

We used the AIA Filament Eruption Catalog (http://aia.cfa.harvard.edu/filament/) \citep{Mc15} as a database for the primary selection of confined filament eruptions. Besides the necessary condition of the absence of the associated CME, we selected events that were located not too close  to the east limb, because there would be difficult to find the initial filament position relative photospheric magnetic fields, and were not too close to the centre of the solar disk, because there would be difficult to estimate the height of the filament above the chromosphere. For the period from May 2013 to July 2014 we selected 16 events listed in Tables 1 and 2.  We begin our selection from most recent events in the catalog, and since eruptive events are rather frequent at solar maximum, all events belong to maximum phase. Whether properties of these failed eruptions differ from failed eruptions in other phases of the solar cycle requires additional examination in future.

In addition to the AIA data we used H$\alpha$ images from the Big Bear Solar Observatory, Kanzelhohe Solar Observatory, and the National Solar Observatory (NSO)/Global Oscillation Network Group (GONG). Observation in EUV with the Sun Earth Connection Coronal and Heliospheric Investigation (SECCHI) Extreme Ultraviolet Imager (EUVI) \citep{Wu04,Ho08} onboard the {\it Solar Terrestrial Relations Observatory} ({\it STEREO}) were very helpful for measurements of eruptive prominence heights and in some cases their initial heights before eruption, if a filament was just at the limb for one of the {\it STEREO} spacecraft. Magnetograms taken by the Heliospheric and Magnetic Imager (HMI) \citep{Sc12}  onboard the {\it SDO} were used as the boundary conditions for the potential magnetic field calculations.

\begin{table*}
\caption{Failed filament eruptions in the quadrupolar magnetic configuration}
\label{T1}
\begin{tabular}{@{}lcccccccccc}
\hline
Date & Coordinates,   & Beginning 
 & X-ray  & Flare & 2a, Mm & $h_0$, Mm 
& $n_0$ & $h_s$, Mm & $n_s$ &  $\alpha_s$, deg
\\
& deg & of the eruption, & flare class &  onset time, & & & & & &
\\
& &  UT &  &  UT & & & & & &
\\
\hline
2013.05.14 & S12 E70 & 05:05 & - & - & 160 & 20 & 0.75
& 150 & 0.5 & 100 \\
2013.06.12 & N33 W10 & 14:45 & - & - & 150 & 20 & 1.25
& 180 & 0.6 & 100  \\
2013.08.04 & N20 E52 & 06:20 & - & - & 120 & 10 & 1
& 120 & 1.2 & 150  \\
2013.12.08 & 00 W90 & 03:15 & - & - & 150 & 40 & 0.8
& 140 & 1 & 80  \\
2014.03.27 & S24 W27 & 12:30 & - & - & 110 & 20 & 1.2
& 220 & 0.8 & 150  \\
2014.03.28 & S10 W66 & 01:30 & - & - & 50 (L) & 15 & 0.75
& 120 & 1.2 & 110  \\
2014.07.02 & N05 W52 & 17:05 & C1.0& 18:00 & 190 & 25 & 1
& 160 & 1.8 & 100  \\
2014.07.05 & S12 W07 & 22:00 & C1.1 & 22:33 & 75 & 25 & 0.7
& 75 & 0.5 & 120  \\
2014.07.30 & S34 E65 & 16:50 & no data & no data & 100 & 20 & 1
& 150 & 0 & 140  \\
\hline
\end{tabular}
\end{table*}

\begin{table*}
\caption{Failed filament eruptions in the dipolar magnetic configuration}
\label{T2}
\begin{tabular}{@{}lcccccccccc}
\hline
Date & Coordinates,   & Beginning 
 & X-ray  & Flare & 2a, Mm & $h_0$, Mm 
& $n_0$ & $h_s$, Mm & $n_s$ &  $\alpha_s$, deg
\\
& deg & of the eruption, & flare class &  onset time, & & & & & &
\\
& &  UT &  &  UT & & & & & &
\\
\hline
2013.05.21 & S15 W62 & 09:45 & C1.2 & 10:23 & 150 & 15 & 0.4
& 150 & 1.8 & 12 \\
2013.06.11 & S07 W42 & 20:20 & - & - & 250 & 40 & 0.6
& 150 & 1.3 & 20  \\
2013.07.03 & N18 W40 & 05:40 & M1.5 & 07:00 & 40 & 10 & 0.4
& 70 & 0.8 & 40  \\
2013.09.20 & S34 W12 & 03:30 & B9.6 & 04:11 & 190 & 30 & 0.7
& 120 & 1.1 & 30  \\
2014.02.17 & S07 W04 & 02:30 & C6.6 & 02:51 & 110 & 20 & 0.8
& 170 & 2.3 & 70  \\
2014.03.20 & N25 W15 & 05:30 & - & -& 110 & 70 & 1.3
& 250 & 2.5 & 25  \\
2014.05.03 & N14 E57 & 07:30 & C1.0 & 07:57 & 25 (L) & 15 & 0.8
& 90 & 2.3 & 5  \\
\hline
\end{tabular}
\end{table*}

Figure 1 shows a typical failed filament eruption observed on 2014 March 28 by  {\it SDO}/AIA in the 304-\AA\ channel close to the west limb (see also Electronic Supplementary Material, Movie1). The filament rises as an arch increasing in size with the legs fixed on the solar surface. The dynamics of the arch summit is presented in the height-time diagram in Fig. 2. The slit used for the construction of the diagram is shown as a white line in Fig. 1e. The filament starts to rise rather abruptly at 13:38 UT with acceleration. After 02:00 UT it begins to decelerate and stops at a height of 120 Mm as a shredded arch. Filament material drains down to the surface along the either arch leg. The arch looks like shrinking a little before disappearance due to draining of the material, however it may be an illusion caused by the falling material. There is no noticeable rotation of the arch plane during the ascent. This is typical for all studied failed eruptions. In the forth column of Tables 1 and 2, the estimated initial filament heights are presented. The accuracy of this estimation is not high because of projection effects and uncertainties of geometrical factors. We will assume an error bar as a half of estimated value, i.e. $\pm$ 20 -- 30\%. This is also true for maximum heights of erupting filaments presented in the sixth column of Tables 1 and 2. The third column presents projected distances $2a$ between the footpoints of arches. These values more or less characterize the real distances along the surface in the case of meridionally oriented filaments and if they are located not far from the solar disc centre. Of course, the projection effect is large for the filament stretched in the west-east direction and located not far from limbs. These cases are marked by L in the tables. The footpoint separations are similar to apex heights of the arches. Therefore, their final shape resembles a crescent. The tables also contain data on the time of the beginning of the filament eruptions and the onset time and class of soft X-ray flares (if any), observed by {\it Geostationary Operational Environmental Satellites} ({\it GOES}) in the long-wavelength band (1 - 8 \AA).

\section{Analysis of the magnetic environment }

\begin{figure*}
		\includegraphics[width=180mm]{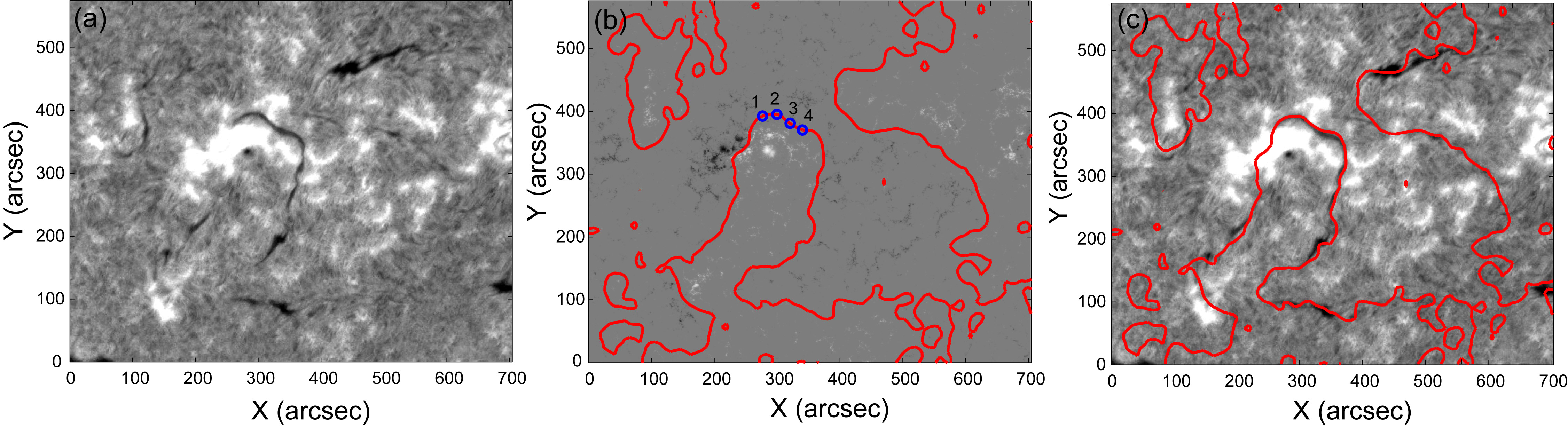}
    \caption{H$\alpha$ filtergram taken on 2014 March 22 at 20:12 UT showing the filament six days before the eruption (a), HMI magnetogram  of the same region and at the same time with superposed PILs (thick red lines) at a height of 6 Mm (b), the same H$\alpha$ filtergram as in the panel (a) with superposed PILs (c). Small blue numerated circles in the panel (b) show the points over which vertical magnetic field profiles were calculated. (Courtesy of the Big Bear Solar Observatory and of the NASA/{\it SDO} HMI science team.)  }
    \label{fig:3}
   \end{figure*}

\begin{figure*}
		\includegraphics[width=150mm]{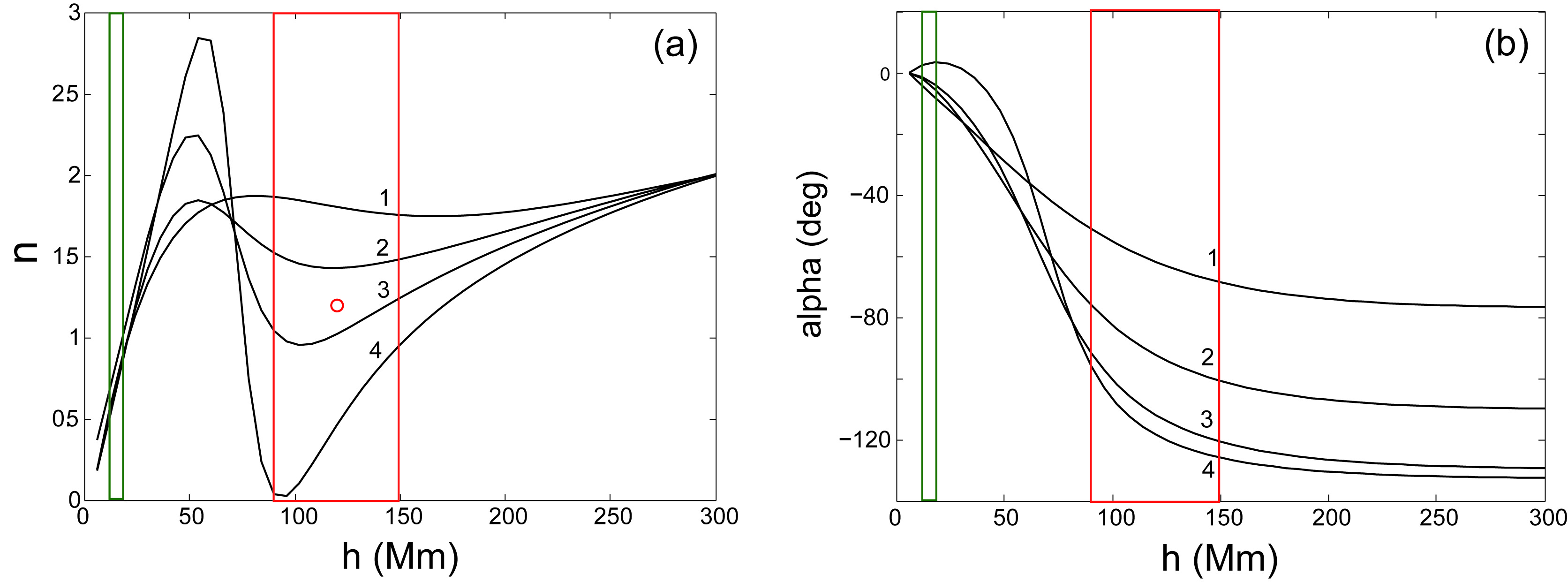}
    \caption{Vertical profiles of the decay index $n$ (a) and the rotation angle $\alpha$ (b) of the horizontal potential magnetic field over the points shown in Fig. 3b. Green and red rectangles show estimated error intervals of initial and final heights of the eruptive prominence, respectively. Small red circle shows the average final height and the value of the decay index. }
    \label{fig:4}
    \end{figure*}

It is widely assumed that coronal magnetic field significantly influences the initiation and prospects of filament eruptions. While coronal electric currents associated with flux ropes and sheared arcades are the source of energy released in eruptions, the field generated by subphotospheric currents determines filament equilibrium conditions and dynamics. In a first approximation, this field can be considered as potential one in the corona. To examine what is specific for the source regions of failed filament eruptions, we calculated the characteristics of potential magnetic field above the filaments. We use the solution of the Neumann external boundary-value problem based on the Green function method (see \citealp{Fi01,Fi13}  and references therein). From a full-disc HMI magnetogram of the longitudinal field we cut a rectangular area around a filament and use it as the boundary condition. Since we selected events remote from the disc centre for the need to observe the shape and height of the eruptive filaments, these areas were not favourable for magnetic-field measurements. We were forced to use  magnetograms taken several days before or after the eruption, when the source region was close to the central meridian and suppose that the large-scale field did not change significantly during this period.

\begin{figure*}
		\includegraphics[width=180mm]{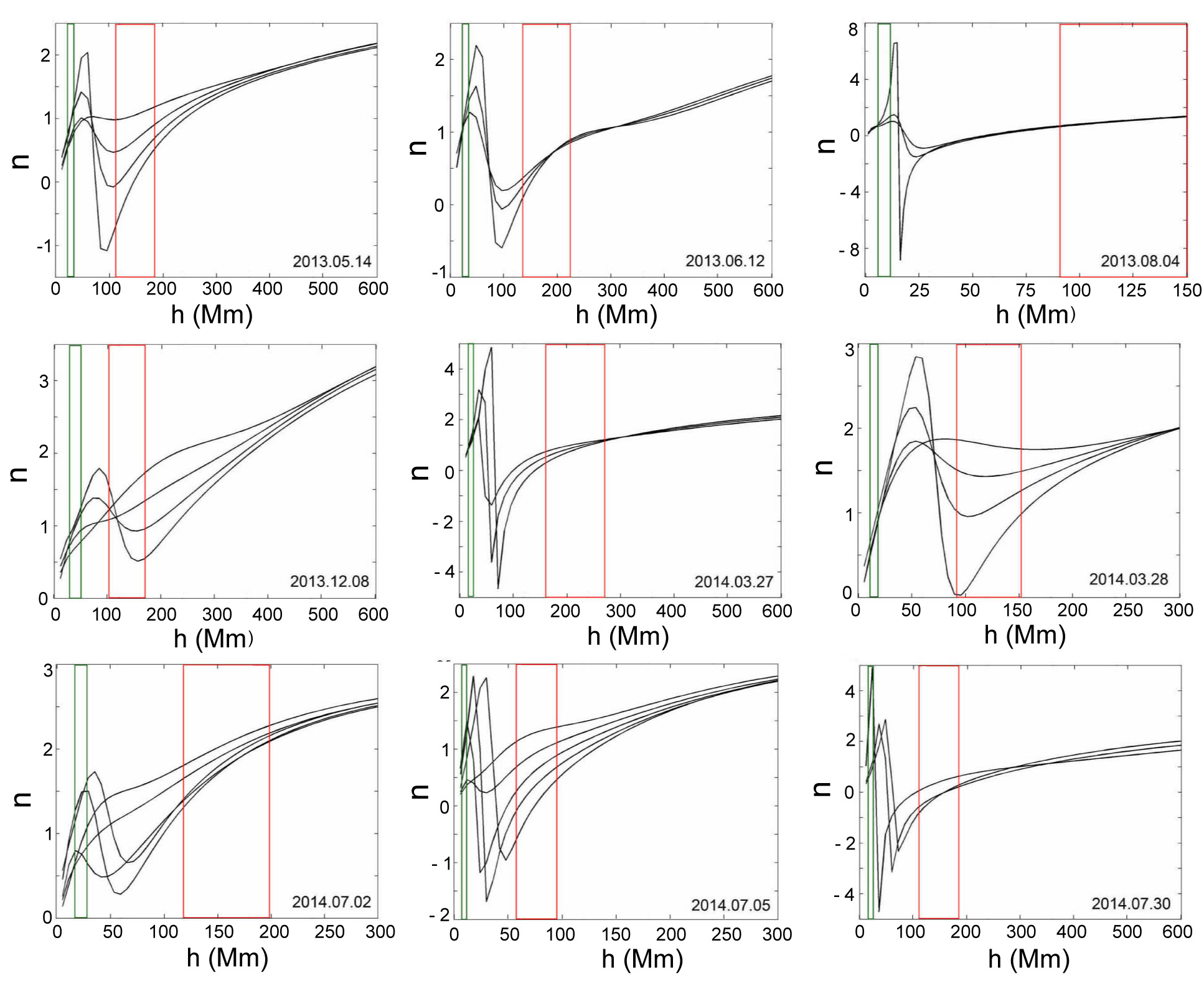}
    \caption{Vertical profiles of the decay index $n$ of potential magnetic field in quadrupolar source regions of failed eruptions.  }
    \label{fig:5}
   \end{figure*}
   
   \begin{figure*}
		\includegraphics[width=180mm]{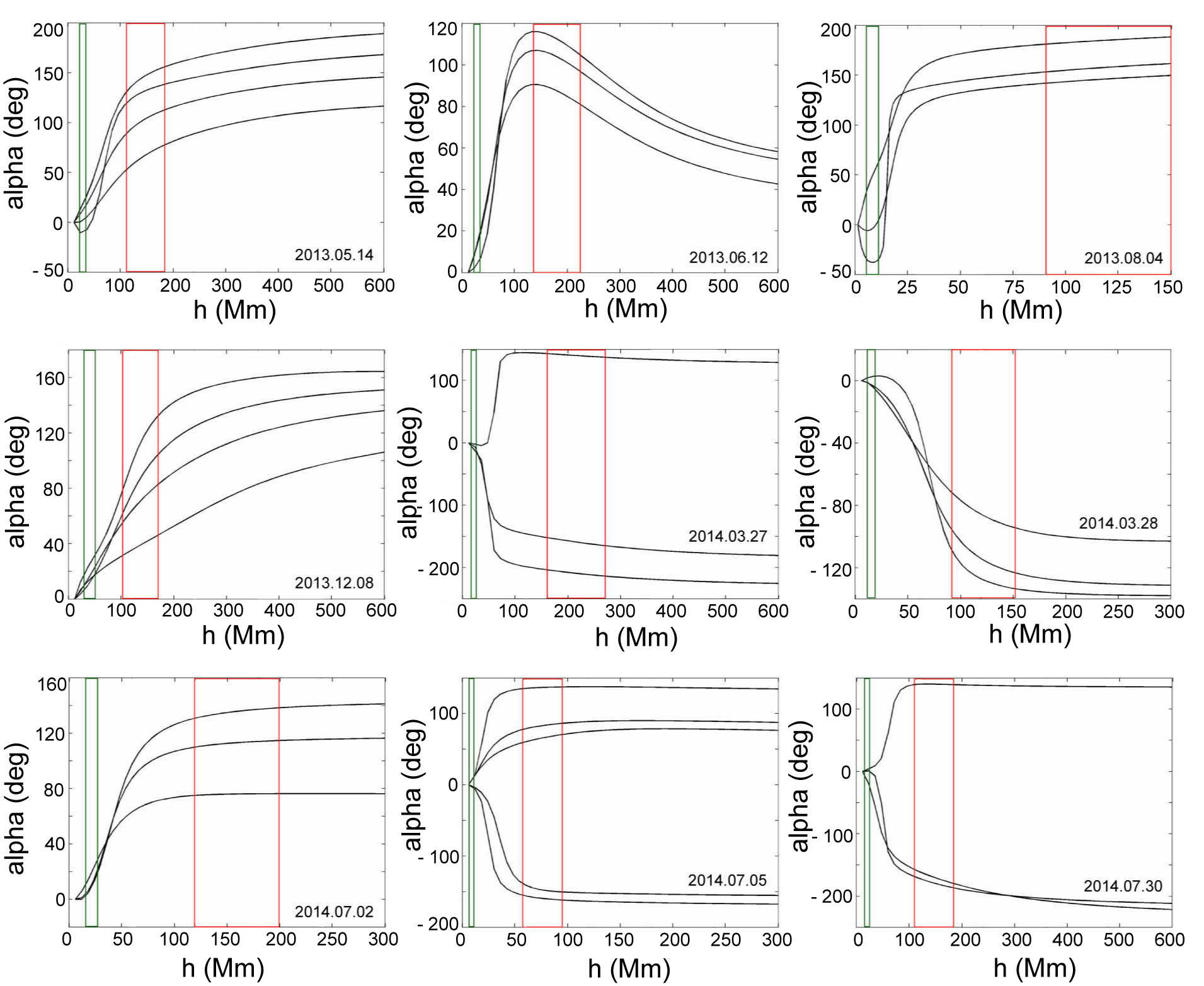}
    \caption{Direction of the potential magnetic field in quadrupolar source regions of failed eruptions. }
    \label{fig:6}
   \end{figure*}

Figure 3a shows the H$\alpha$ filtergram of the source region of the eruption presented in Fig. 1 six days before the event. The eruptive filament represents the north-east section of a rather long filament. Figure 3b demonstrates the HMI magnetogram of the same region at the same time. Both images are slightly remapped in order to obtain an array with pixels of equal area more suitable for the magnetic extrapolation  \citep{Fi13,Su13}. The red contours present the polarity inversion line (PIL) at the height of 6 Mm superposed on the magnetogram (Fig. 3b) and the H$\alpha$ filtergram (Fig. 3c). Small blue numerated circles show the points over which vertical magnetic field profiles were calculated. 
  
Figure 4 shows vertical profiles of the decay index $n$ of the potential coronal magnetic field calculated using Equation (1). The profiles become steeper curved as switchbacks closer to the western end of the filament section that later erupted.\citet{Wa17} called such type of decay index profiles as saddle-like ones. The value of the decay index at the initial height of the filament before the eruption (within the error bar indicated by the green rectangle) is close to unity. The maximum height lies within the error bar shown by the red rectangle. While there is a wide range of $n$ within the red rectangle, an average value, which possibly corresponds just to the summit of the eruptive-filament arch, is between 1 and 1.5. The steep changes of the decay index suggest the proximity to a magnetic null point. Indeed, the horizontal field turns steeply at the same heights and becomes nearly oppositely directed (Fig. 4b). At the maximum height of the eruptive filament, the horizontal field rotates through about 100$^\circ$ from its near-surface direction. These properties of the coronal potential field are evidence of a quadrupolar magnetic configuration in the source region. 

\begin{figure*}
		\includegraphics[width=180mm]{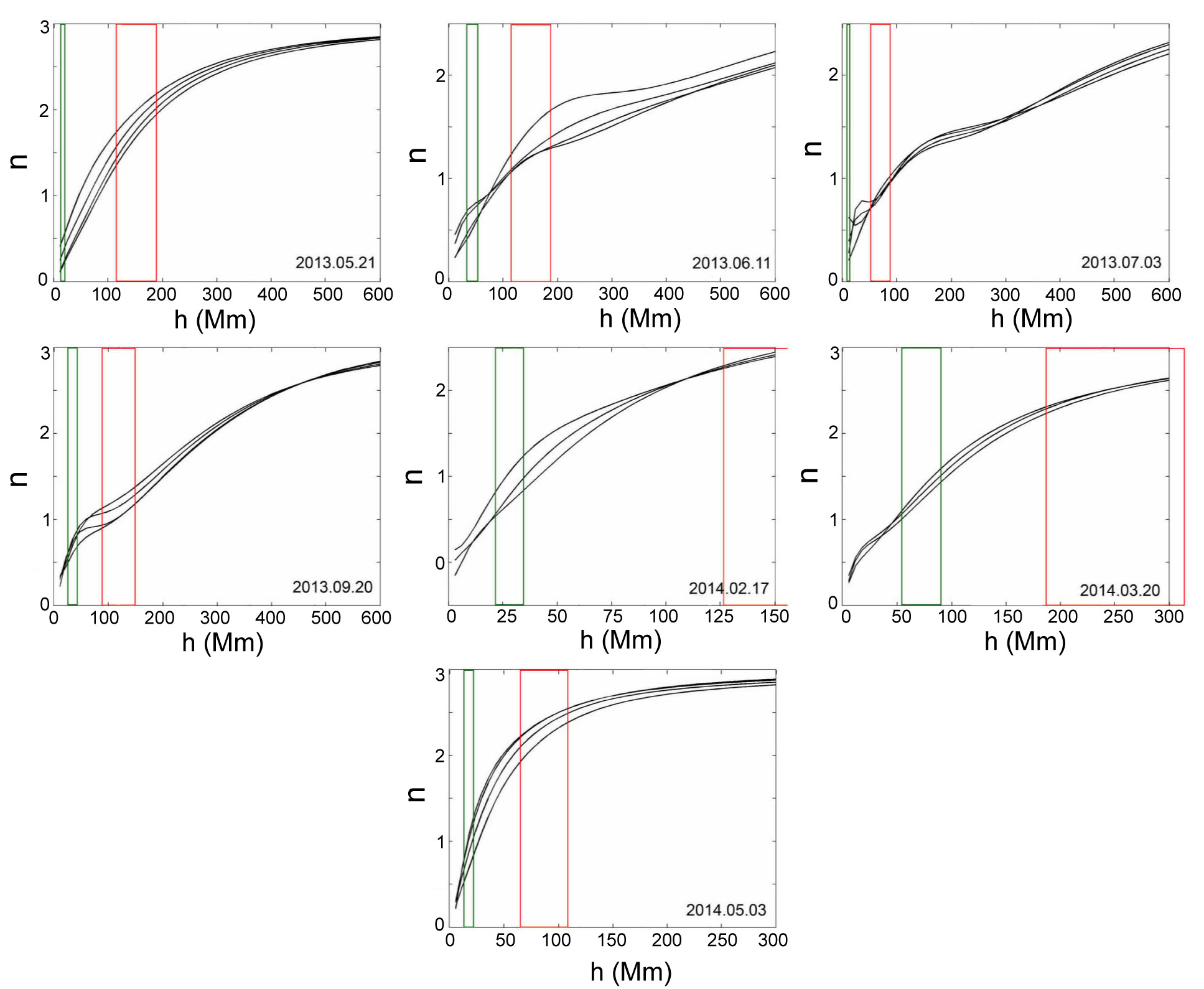}
    \caption{Vertical profiles of the decay index  $n$ of potential magnetic field in dipolar source regions of failed eruptions.  }
    \label{fig:7}
   \end{figure*}
   
   \begin{figure*}
		\includegraphics[width=180mm]{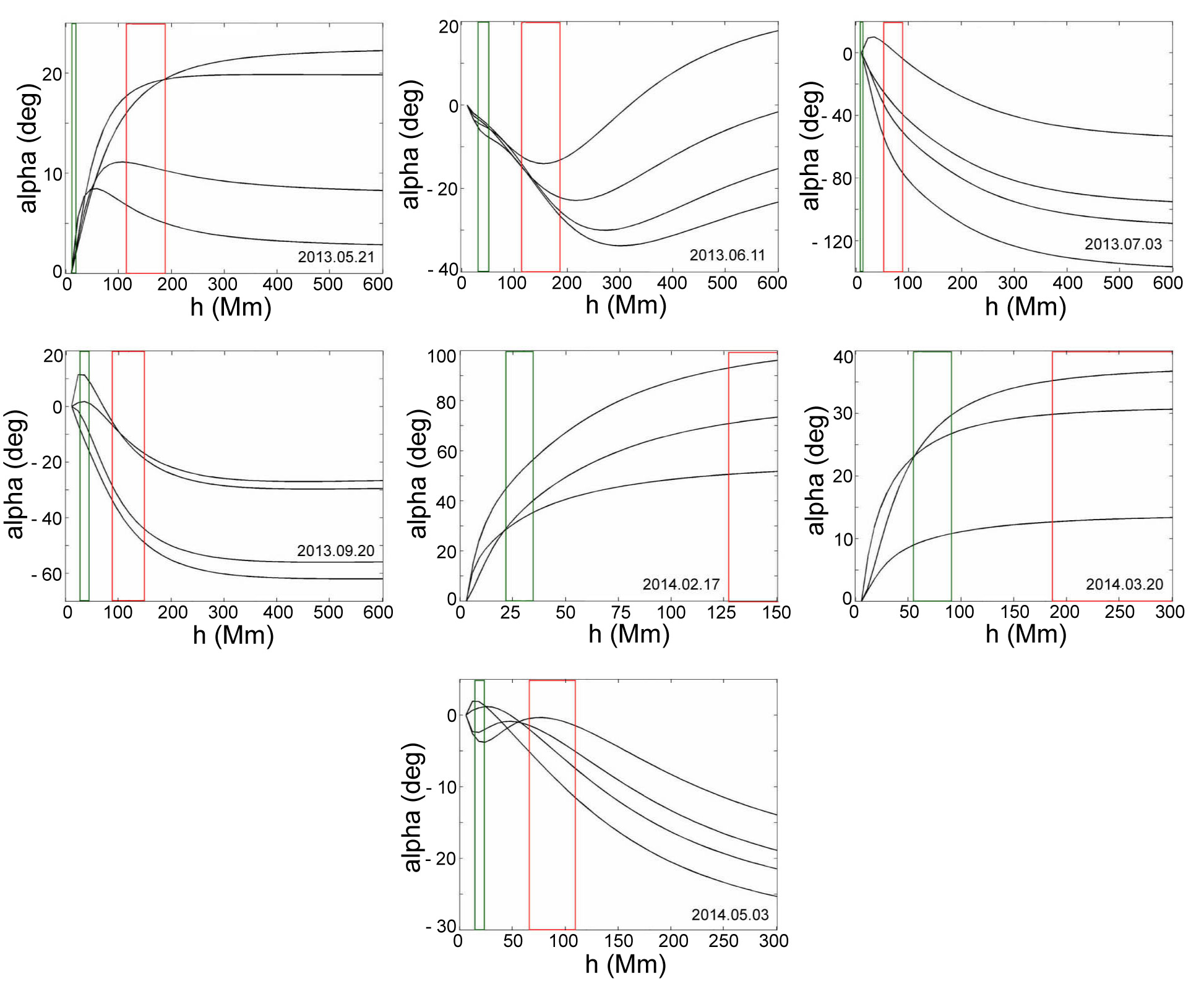}
    \caption{Direction of the potential magnetic field in dipolar source regions of failed eruptions. }
    \label{fig:8}
   \end{figure*}
   
The quadrupolar magnetic configuration is found in more than half of studied events. Figures 5 and 6 present the decay index $n$ and the rotation angle $\alpha$ profiles analogous to the profiles shown in Fig. 4. These events are listed in Table 1. The decay index at the initial height in these eruptions is close to unity with the dispersion of about $\pm$\,0.25 (the fifth column of Table 1). The filaments stop at heights above null points where the decay index is increasing with height but not too great being in the range from 0 to 1.8 (the seventh column of Table 1). The rotation angles $\alpha$ presented in the eight column are more than 90$^\circ$ at the filament maximum height in all cases except one.  

The other events (7 from 16) originate from source regions with a different structure. The decay index in these regions increases monotonically with height (Fig. 7). The magnetic structure can be assumed as dipolar. There are some deviations from a smooth curve showing the presence of different scales in the magnetic field but they do not significantly violate the general behaviour of $n$. The rotation of the field does not exceed 70$^\circ$ at filament maximum heights, being usually much lesser. These events are listed in Table 2. These filaments start to accelerate at heights with decay index values in the range from 0.4 to 0.8 except one case with $n$ = 1.2 (the fifth column of Table 2). In contrast, the decay index at maximum heights is much greater. It is in the range from 1.1 to 2.5, also except one case with $n$ = 0.8 (the seventh column of Table 2).

\section{Magnetic flux-rope equilibrium and stability}

\begin{figure}
		\includegraphics[width=90mm]{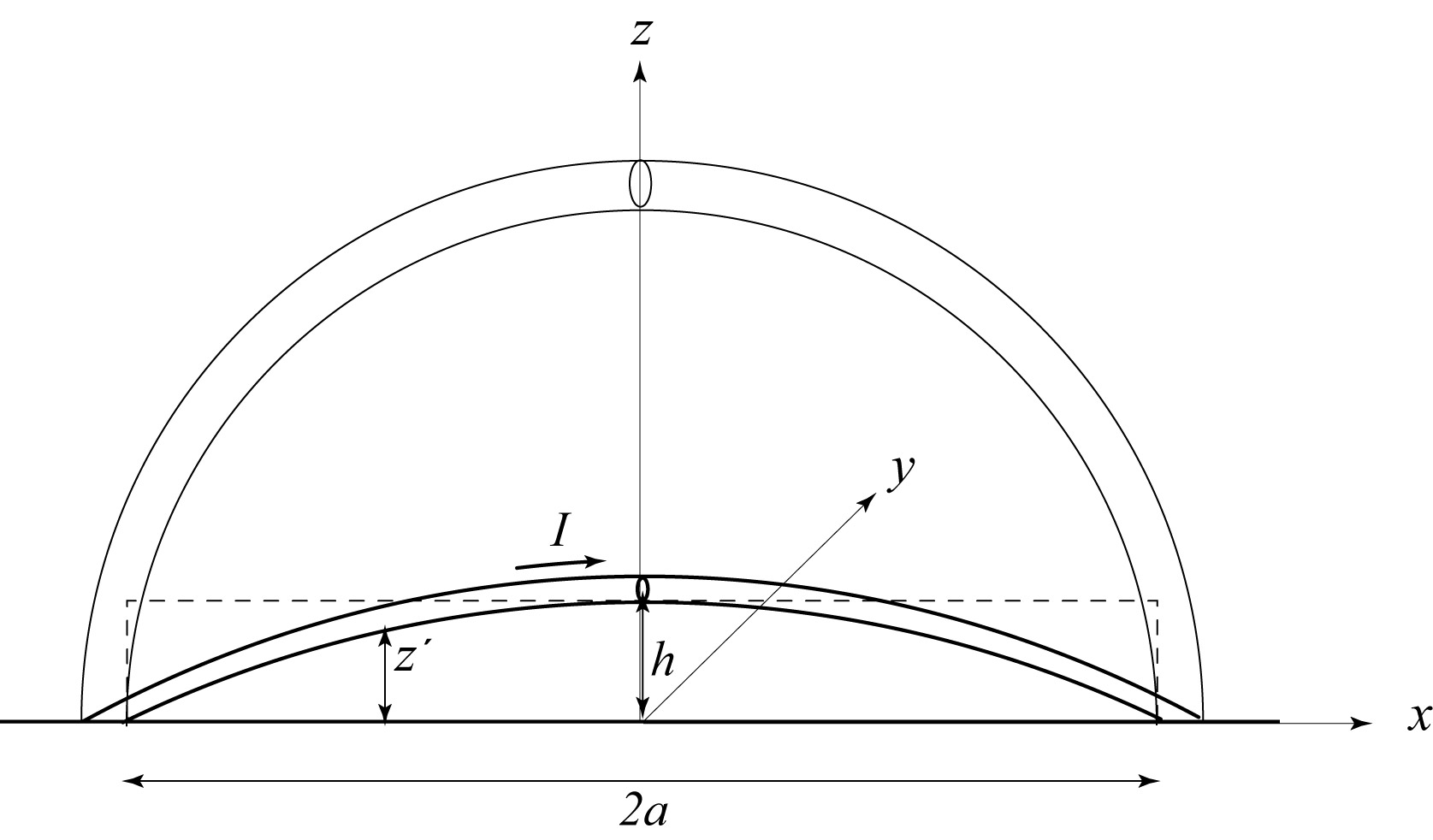}
    \caption{Schematic drawing of an erupting circular flux rope}
    \label{fig:9}
\end{figure}

In all studied events, the filament eruptions start at a height where the decay index is about unity or less. In models of thin flux-rope equilibrium, this value of the decay index is typical for the loss of equilibrium of a flux rope with the straight axis \citep{va78,Mo87,De10}. More than half of the filaments in the dipolar magnetic configuration stop at a height where the decay index is greater than 1.5, the value considered as the threshold of the torus instability for thin flux ropes with the circular axis \citep{Ba78,Kl06}. In the quadrupolar source regions, the filaments stop above null points at a height where the decay index is less than unity which at first glance is favourable for a stable equilibrium, but the horizontal coronal filed rotates there through the angle more than 90$^\circ$ from its near-surface direction and cannot more hold the flux rope.

In order to find what could be the reason of the termination of the filament ascent, we analyse a simple model of the thin flux-rope equilibrium. We consider the flux rope as a section of a torus with its two footpoints anchored in the photosphere as shown in Fig. 9 \citep{Ca94,Ol10}. The initial major radius of the torus $R$ is large because the initial height of the flux-rope apex $h$ is low
\begin{equation}
R =  \frac{a^2 + h^2}{2h} ,
\end{equation}  
where $a$ is the half-footpoint separation. In neglect of plasma pressure forces in comparison with the Lorenz forces (low-beta plasma), we consider the equilibrium of the flux-rope apex as a balance of the following forces.
The upward Lorenz self-force (per unit length), also referred as the hoop force, strongly depends on the curvature radius \citep{Sh66} 
\begin{equation}
F_h =  \frac{I^2}{c^2R} \left(\ln \frac{8R}{r} - 1 + \frac{l_i}{2}\right) ,
\end{equation}  
where $I$ is the toroidal electric current, $r$ is the minor radius of the torus, $l_i$ is the internal inductance per unit length. This force vanishes at low heights, since the axis curvature becomes large, and nearly straight electric current does not influence itself. On the other hand, the low flux rope is under the strong influence of the diamagnetic photosphere creating the upward force usually modelled by acting of the mirror current \citep{Ku74} 
\begin{equation}
F_I =  \frac{I^2}{c^2h}.
\end{equation}  

Equations (3) and (4) describe the limiting cases of the force acting on the flux-rope element from the whole electric circuit. $F_I$ corresponds to the case of a low height, when the influence of the inductive surface-currents is maximal, while $F_h$ is negligible. In contrast, $F_h$ describes the action of neighbour flux-rope sections when the surface currents are at a greater distance and their action is unimportant. The sum of $F_I$ and $F_h$ can be approximately assumed as the expression valid at any height. The calculation of the precise expression is a challenging task \citep{Ga94}.

The internal toroidal magnetic field $B_t$ creates the downward tension force
\begin{equation}
F_t = - \frac{\bar{B_t^2} r^2}{8R} ,
\end{equation}  
where the bar denotes the averaging over the minor radius $r$, while the external constraining poloidal magnetic field $B_e$ acts downwards with the force 
\begin{equation}
F_B =  -\frac{IB_e}{c} .
\end{equation}  
We also take into account the gravity force $F_g = Mg$, where $M$ is the mass of the filament per unit length and $g$ is the free-fall acceleration, although this force is often considered negligible at low heights compared with the Lorenz forces  \citep{Mo87,Pr90}. In the interval of heights typical for failed eruptions, $g$ can be considered as a constant.

Equilibrium along the minor radius in neglect of plasma pressure gradient demands the following relationship between toroidal $B_t$ and poloidal $B_p$ components of the internal magnetic field \citep{Sh66} 
\begin{equation}
\bar{B_t^2} =  B_p^2 ,
\end{equation}  
where 
\begin{equation}
B_p = - \frac{2I}{cr} .
\end{equation}  
The condition (7) is satisfied for force-free magnetic field. Following \citet{Li98} and  \citet{Is07}  we chose the linear force-free internal magnetic structure \citep{Lu51}, which gives $l_i = 1$. Then taking into account Equations (3), (5), (7), and (8) the radial self-force per unit length is given by 
\begin{equation}
F_R =  F_h + F_t =\frac{I^2}{c^2R} \left(\ln \frac{8R}{r} - 1 \right) .
\end{equation}  
Conservation of the toroidal flux within the flux rope with the linear force-free field determines the dependence of the flux-rope radius $r$ on the current $I$ \citep{Li98}:
\begin{equation}
r =  r_0\frac{I_0}{I} .
\end{equation}  

Changes of the electric current $I$ during eruptions are usually connected with the conservation of the poloidal magnetic flux between the photosphere and the flux rope:  
\begin{equation}
\Phi_p =  \Phi_I + \Phi_s = cLI + \int_sB_e ds = \mbox{const},
\end{equation}  
where $S$ is the area between the photosphere and the flux rope and $L$ is the self-inductance of the torus fraction above the photosphere \citep{Ca94,Ol10}:
\begin{equation}
L = \frac{4\pi R \Theta}{c^2} \left(\ln{ \frac{8R}{r}} - \frac{3}{2} \right) ,
\end{equation}  
with the $\Theta$-factor indicating the portion of a circle above the photosphere:
\begin{equation}
 \Theta  = \left\{ \begin{array}{ll}
 1 - \frac{\theta}{\pi} \:, &  h \geq a \;,
\\ & \\ 
\frac{\theta}{\pi}  \:, & h < a \;,
\end{array} \right.
\end{equation}
and  							\begin{equation}
\theta =  \sin^{-1}{\frac{a}{R}} .
\end{equation}  

Since $L$ depends on $r$, while $r$ depends on $I$, in calculating $I$ at some height we need to use at first the initial value of $r$, $r = r_0$, then to substitute the value of $I$ into Equation (10) and use the new value of $r$ for the calculation of $I$. After several iterations we obtain self-consistent value of $I$.

For the external field we chose a simple presentation as the field of two 2D horizontal dipoles $m_1$ and $m_2$ located at different depths $d_1$ and $d_2$ below the conductive surface (photosphere): 
\begin{equation}
B_e =  \frac{m_1}{(z + d_1)^2} + \frac{m_2}{(z + d_2)^2} ,
\end{equation}  
where $z$ is the vertical coordinate. The case of one dipole was studied in detail for 2D model by \citet{Pr90}. Combination of fields of two dipoles allows us to have various vertical profiles of the external field. \citet{Fi18} showed that two dipoles with the same directions but with different dipolar moments and at different depths are able to create an additional equilibrium point in the corona, which may serve as a maximum height for a failed eruption or an intermediate metastable state in a two-step eruption of a filament.  

Despite the simplicity of the expression for $B_e$, the exact value of the flux $\Phi_s$ can be calculated only numerically, e.g. for $h < a$:
\begin{equation}
\Phi_s(a, h) =    \int_{-a}^a\int_0^{z^{\prime}} B_e dx dz = \int_{-a}^a\left(\frac{m_1}{(z^{\prime} + d_1)} + \frac{m_2}{(z^{\prime} + d_2)} \right) dx ,
\end{equation}  
where $z^{\prime}$ is the $z$-coordinate of the circular contour (Fig. 8): 
\begin{equation}
z^{\prime} =  \frac{\left[(a^2 + h^2)^2 - 4 h^2 x^2 \right]^{1/2} - a^2 + h^2}{2 h} .
\end{equation}  
 However, calculations show that the value of the flux below the circular contour is very close to the value of the flux through the rectangular contour shown in Fig. 9, which can be expressed as
 \begin{equation}
\Phi_s(a, h) \approx 2a \left(\frac{m_1}{(h + d_1)} + \frac{m_2}{(h+ d_2)} \right) .
\end{equation}  

\begin{figure}
		\includegraphics[width=70mm]{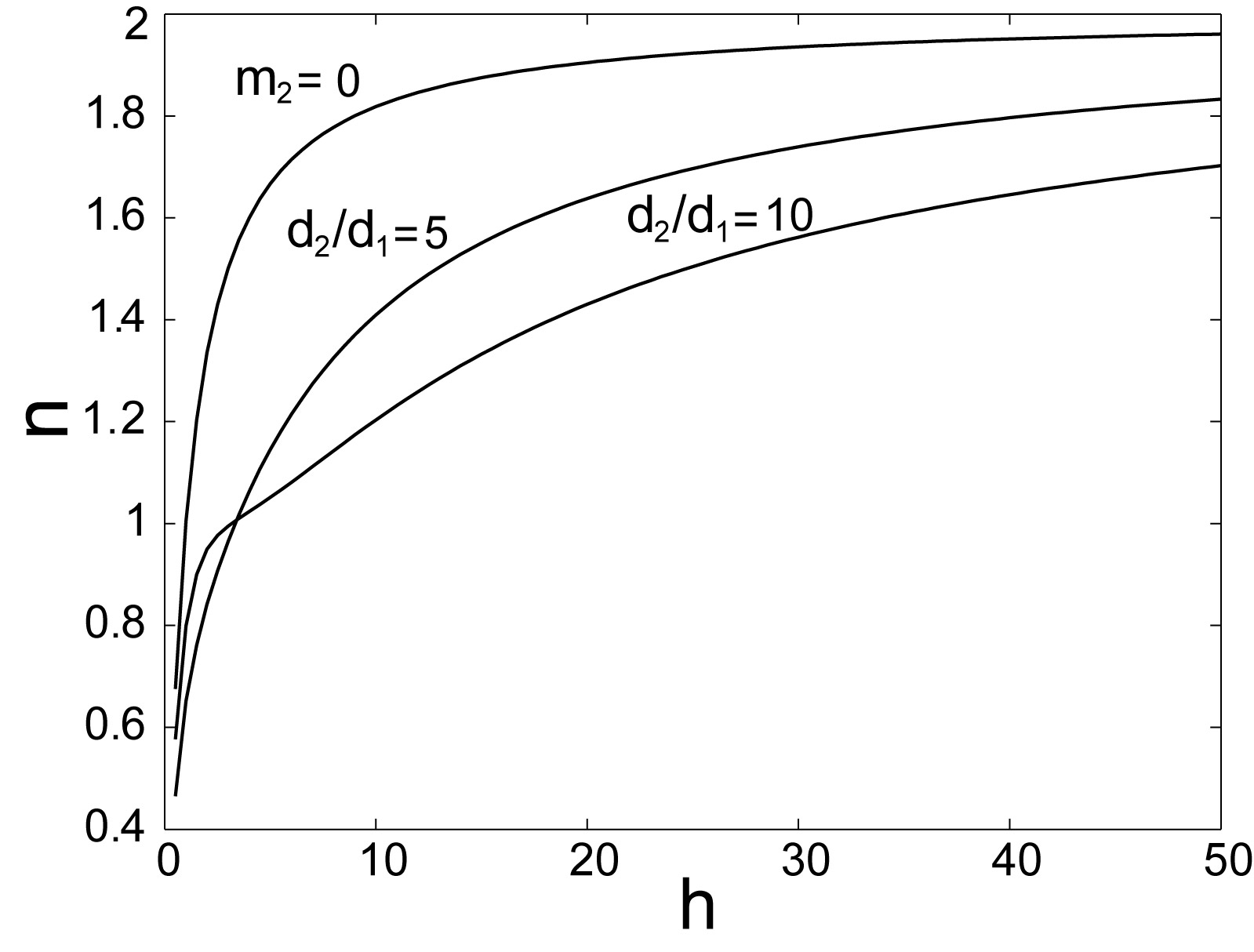}
    \caption{Vertical profiles of the decay index n of the field of a single dipole and two dipoles with $m_2/m_1 = 10$ and different depths. The horizontal axis is in units of $d_1$.}
    \label{fig:10}
\end{figure}

Let us consider at first the case when the vertical decay index profile is monotonic as in Fig. 7, which requires that either dipole has the same direction.  The field is perpendicular to the plane of the flux-rope loop and has the same direction everywhere. We may expect that in contrast to the switchback $n$-profile used by \citet{Fi18} to model a two-step eruption, there could be the second equilibrium point at greater height than initial equilibrium with $1 < n < 1.5$, where a curved flux rope could find the stable equilibrium. Figure 10 shows $n$-profiles of a single 2D dipole and two dipoles with $m_2/m_1 = 10$ and different depths of the greater dipole. The decay index of the field of the single dipole increases rather steep, while with the ratio of depth of two dipoles of the order of 10, the profile is gently sloping but still monotonic. We use this profile for the analysis of equilibrium (Fig. 11). We should specify then the footpoint separation. A great value of $a$, compared to $d_1$ used as a scale in our model, leads to a long extent of the radial self-force $F_R$ increase. Too short separation means a curved initial axis of the flux rope. We chose $a = 3d_1$, so that the flux rope will be straight enough to have the critical decay index close to unity.

Figure 11b demonstrates the behaviour of the forces, acting on the flux rope according Equations (6), (8), and (9), with height in units of $d_1$. The gravity force is chosen to be of the order of one tenth of the Lorenz forces at the initial equilibrium point (the green line). The total force $F$ acting on the flux rope is presented by the thick red line. The initial value of the electric current $I$ is chosen to have an unstable equilibrium point at a low height, which is the start point for the eruption. The red line touches the horizontal axis showing the point of unstable equilibrium. This point is located at the height lower than the value of $d_1$ and corresponds to the value of the decay index less than unity. This is apparently due to the increase of the electric current $I$ according the inductance equation (11) at low heights (Fig. 11c), because of the rather steep increase of $\Phi_s$ and slight changes of $L$ at low heights. 

While $F_I$ and $F_B$ rapidly decrease with height, the radial self-force $F_R$ at first increases, as the curvature of the flux rope increases according Equation (2), until $h = a$ and then decreases, as the curvature decreases. The total force $F$ vanishes at $h \approx 9$ and becomes directed downward at greater heights. Therefore, any displacement of the flux rope away from this point restores it back: the total force is directed upward below the point and downward above it. Hence, this is stable equilibrium point, which can be considered as the maximum height of the flux rope in the failed eruption, if some drag force prevents its oscillations \citep{Fi18,Za18}. Obviously that this equilibrium is reached due to the presence of the gravity force. Without gravity, the total force slowly decreases but is hold directed upwards like the red curve approaches to the green line in Fig. 11b. In principle, the upper equilibrium point can exist without taking into account the gravity force in much slower decreasing magnetic field but anyway it would be at rather great height and the $n$-profile would not be very similar to the profiles shown in Fig. 7.

\begin{figure*}
		\includegraphics[width=180mm]{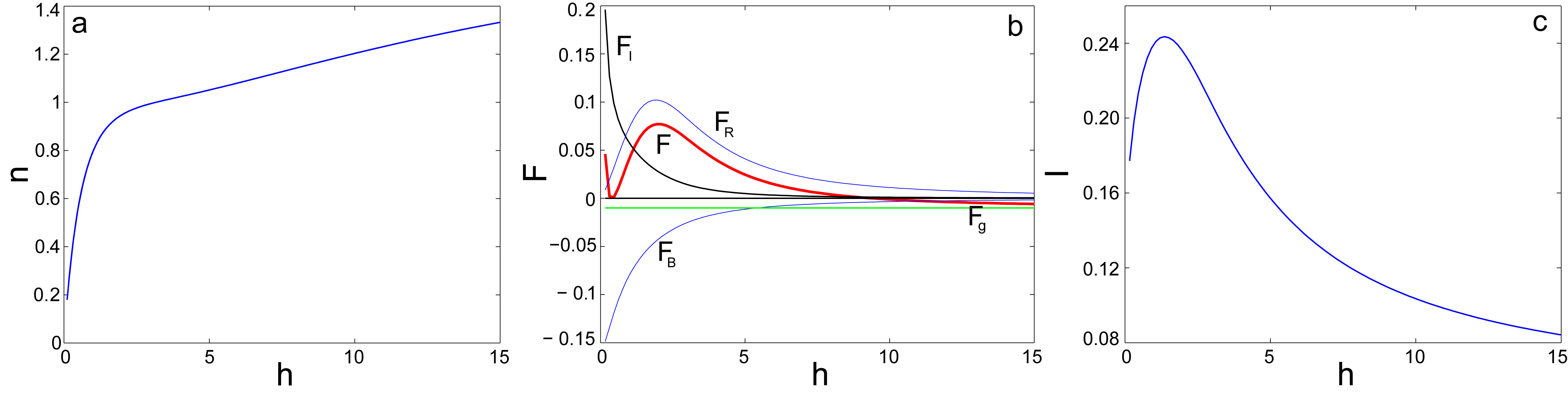}
    \caption{Vertical distributions of the decay index $n$ (a), forces acting on the flux rope (b), and the value of the electric current $I$ (c) in the dipolar external field. $m_2/m_1 = 10, d_2/d_1 = 10$, horizontal axes are in units of $d_1$.   $F$ and $I$ are in dimensionless units for $d_1 =1, m_1 = 1$.}
    \label{fig:11}
   \end{figure*}
   
   \begin{figure*}
		\includegraphics[width=180mm]{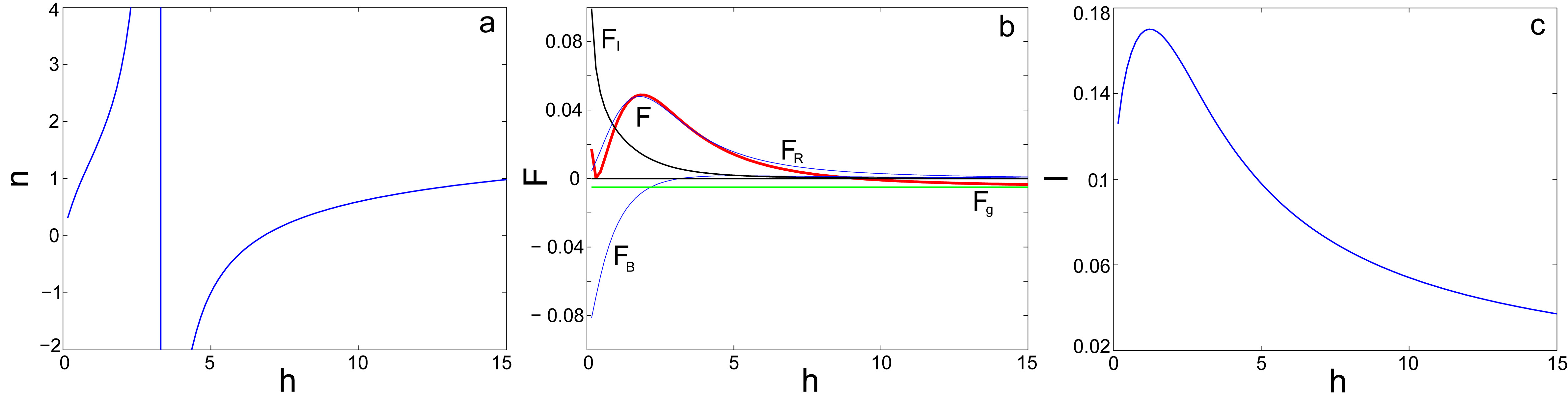}
    \caption{Vertical distributions of the decay index  $n$ (a), forces acting on the flux rope (b), and the value of the electric current $I$ (c) in the quadrupolar external field. $m_2/m_1 = -10, d_2/d_1 = 10$, horizontal axes are in units of $d_1$.  $F$ and $I$ are in dimensionless units for $d_1 =1, m_1 = 1$.}
    \label{fig:12}
   \end{figure*}

In the quadrupolar magnetic configuration, without taking into account the gravity force there is no chance for the flux rope to find a new equilibrium above a null point, where the external field does not hold it more but accelerates the flux rope upwards. The field is perpendicular to the plane of the flux-rope loop and change direction to the opposite above the null point. Figure 12 shows the height-dependence of values in the quadrupolar configuration similar to presented in Fig. 5. The decay-index jump at $h \approx 3.5$ (Fig. 12a) corresponds to the null point above which the field changes direction to opposite. The force $F_B$ is positive (upward) above the null point as well as the other forces action on the flux rope except the gravity force (Fig. 12 b). In general, the pattern is very similar to presented in Fig. 11. The point of instability is also located at the height lower than the value of $d_1$ and corresponds to the value of the decay index less than unity. 

\section {Discussion and conclusions}

We analyze 16 failed filament eruptions observed near 24 solar maximum from May 2013 to July 2014. Besides the necessary condition of the absence of the associated CME, we selected events that were not located too close the east limb, because there would be difficult to find the initial filament position relative photospheric magnetic fields, and were not too close to the centre of the solar disk, because there would be difficult to estimate the height of the filament above the chromosphere. In contrast to events studied by \citet{Zh19}, no significant rotation of the filament arch plane is observed during the ascent in all studied failed eruptions. Therefore, there is no evidence of writhing and kink-instability in this sample of events, although we cannot rule out absolutely the kink-instability as a trigger of eruptions.

Equilibrium and dynamics of a magnetic flux rope in the corona depends on the external magnetic field generated mostly by electric currents located below the photosphere. Potential field approximation is widely used as a rather simple technique to extrapolate the photospheric fields into the corona. We calculated potential magnetic field distributions in the corona above the initial locations of the filaments using in some cases magnetograms several days before or after eruptions when the source region was close to the central meridian. The large-scale coronal magnetic field is supposed to have minor changes during this period of time. In seven events, the height-dependence of the decay index is monotonic. The filaments start to erupt at a height where the decay index is a little lesser then unity and stop at a height where $n > 1.5$ for four events. These events should be torus-unstable at the maximum height and there must be another cause apart from the strapping coronal field to confine them.  

The other nine events occur in the regions with the switchback or saddle-like $n$-profiles. At first glance, this shape of the profiles is favourable for the existence of a torus-stable zone ($n < 1.5$) high in the corona \citep{Wa17,Fi18,Liu18}  because after the initial increase the decay index drops to values below the threshold of the torus instability. On the other hand, it was these regions with the saddle-like $n$-dependence that were eruptive in the sample studied by \citet{Ch11}. \citet{In18} on the basis of the numerical MHD simulation showed that that an eruptive flux rope can pass through a torus-stable zone ($n < 1.5$) high in the corona in a region  with the saddle-like vertical decay index profile because of a nonlinear positive feedback process between the flux rope and magnetic reconnection behind it. The coronal magnetic field in their simulation was calculated using the measured photospheric magnetogram partly in the nonlinear-force-free field approximation and partly in the current-free approximation. It is not evident from the paper whether the horizontal field changes direction at the bottom of the saddle-like $n$-profile. In all our events, the direction of the horizontal field near the saddle bottom is turned through more than 100$^\circ$ relative its direction at the initial filament position. Since the apex of the eruptive filaments does not rotate significantly in our events, the external field does not hold it more but accelerates the flux rope upwards. 

In order to find what could be the reason of the termination of the filament ascent, we analyse a simple model of the thin flux-rope equilibrium. The external magnetic field is modelled by the field of two linear dipoles with different dipole moments and located at different depths below the photosphere. This simple model allows us to reproduce different vertical behaviour of the external field. The flux rope is considered as a section of a torus with its two footpoints anchored in the photosphere. The initial major radius of the torus is large because the initial height of the flux-rope apex is low. That is why the initial equilibrium of the flux rope is similar to the equilibrium of a straight linear current, stable if the external magnetic field decreases with height not too fast ($n < 1$). However, the exact value of the critical value of the decay index depends also (apart from some other factors that we do not discuss here) on the variations of the electric current with height according the inductance equation. At low heights, the current value increases due to increase of the magnetic flux of the background field through the flux-rope contour. As a result, the critical value of the decay index becomes smaller than unity. In the majority of the studied events, the eruptions really start from the heights where the decay index is smaller than unity. 

When the height of the flux-rope apex increases, its axis becomes more curved leading to the increase of the hoop force. After the reaching the maximum value at the height equal to half of the footpoint separation, the hoop force slowly decreases. The need of the rather steep hoop-force decrease limits the footpoint separation. 

In a pure 2D model, the confinement of an eruption is possible only in the external magnetic field with the saddle-like $n$-profile and not changing the direction \citep{Fi18}. There are no events in our sample with similar properties of the coronal field. In a 3D model with a curved flux rope, the existence of the second stable equilibrium point higher in the corona is, in principle, possible even in a region with the monotonic $n$-profile. The eruption can start at a low height where the flux-rope curvature is small and $n_c \approx 1$ and be confined at a greater height where $n \leq 1.5$, since the flux rope becomes circular with the threshold for the torus instability $n_c = 1.5$. However, the $n$-profile should be rather flat and, anyway, the possible upper equilibrium-point position is very high compared to the starting point.  

In the quadrupolar source regions with the switchback decay-index profiles, all components of the Lorenz force are directed upward above a null point. The most reasonable force that can terminate ascending of the flux rope and balance the Lorenz force seems gravity. The gravity force can be negligible at a low height compared with any component of the Lorenz force and not influence on initial equilibrium and stability of the flux rope, although under certain condition it can increase the value of $n_c$ \citep{Ts19}. The situation is quite different at greater heights. In the height interval of interest ($h$\,<\,200 Mm), the free-fall acceleration can be considered as constant, so the gravity force is also constant, if the draining of the material from the filament is negligible. In contrast, all components of the Lorenz force decrease rapidly with height. At some height the gravity force balances the upward forces and maintains the stability. Of course, the shape of the flux rope becomes more or less circular, and the filament material is able to drain down along the sloping legs of the flux rope, that is really observed. But this process is rather slow in comparison with the eruption. In fact, filament arches are not observed for a long time after failed eruptions, but loose their material through the draining and fade out on the background coronal emission, The remained  magnetic structure of the flux rope can ascend slowly after the lost of material and merge with the coronal magnetic field.

We expected that the structure of the coronal magnetic field would be decisive for the prospects of a filament eruption, whether it would be successful or failed. However, in the studied sample of failed eruptions we did not find in the structure of the coronal field any specific features that can be considered as responsible for confinement of eruptions. We come to conclusion that in these limited-size events the initial mass of filaments and the corresponding gravity force determine the confinement.

\section*{Acknowledgements}

The author thanks the Big Bear Solar Observatory, Kanzelhohe Solar Observatory, the National Solar Observatory/Global Oscillation Network Group, the {\it SOHO}, the {\it STEREO}, and the {\it SDO} scientific teams for the high-quality data they supply. The author thanks the referee for critical comments and useful suggestions. {\it SOHO} is a project of international cooperation between ESA and NASA. {\it STEREO} is the third mission in NASAs Solar Terrestrial Probes program. {\it SDO} is a mission of NASAs Living With a Star Program. {\it GOES} is a joint effort of NASA and the National Oceanic and Atmospheric Administration (NOAA). The movie was created using the ESA and NASA funded Helioviewer Project.




\bibliographystyle{mnras}
\bibliography{Reference} 

\begin{thebibliography}{}
\makeatletter
\relax
\def\mn@urlcharsother{\let\do\@makeother \do\$\do\&\do\#\do\^\do\_\do\%\do\~}
\def\mn@doi{\begingroup\mn@urlcharsother \@ifnextchar [ {\mn@doi@}
  {\mn@doi@[]}}
\def\mn@doi@[#1]#2{\def\@tempa{#1}\ifx\@tempa\@empty \href
  {http://dx.doi.org/#2} {doi:#2}\else \href {http://dx.doi.org/#2} {#1}\fi
  \endgroup}
\def\mn@eprint#1#2{\mn@eprint@#1:#2::\@nil}
\def\mn@eprint@arXiv#1{\href {http://arxiv.org/abs/#1} {{\tt arXiv:#1}}}
\def\mn@eprint@dblp#1{\href {http://dblp.uni-trier.de/rec/bibtex/#1.xml}
  {dblp:#1}}
\def\mn@eprint@#1:#2:#3:#4\@nil{\def\@tempa {#1}\def\@tempb {#2}\def\@tempc
  {#3}\ifx \@tempc \@empty \let \@tempc \@tempb \let \@tempb \@tempa \fi \ifx
  \@tempb \@empty \def\@tempb {arXiv}\fi \@ifundefined
  {mn@eprint@\@tempb}{\@tempb:\@tempc}{\expandafter \expandafter \csname
  mn@eprint@\@tempb\endcsname \expandafter{\@tempc}}}

\bibitem[\protect\citeauthoryear{{Alexander}, {Liu}  \& {Gilbert}}{{Alexander}
  et~al.}{2006}]{Al06}
{Alexander} D.,  {Liu} R.,   {Gilbert} H.~R.,  2006, \mn@doi [\apj]
  {10.1086/508137}, \href {http://adsabs.harvard.edu/abs/2006ApJ...653..719A}
  {653, 719}

\bibitem[\protect\citeauthoryear{{Bateman}}{{Bateman}}{1978}]{Ba78}
{Bateman} G.,  1978, {MHD instabilities}.
MIT Press, Cambridge, Mass.

\bibitem[\protect\citeauthoryear{Baumgartner, Thalmann  \& Veronig}{Baumgartner
  et~al.}{2018}]{ba18}
Baumgartner C.,  Thalmann J.~K.,   Veronig A.~M.,  2018, \mn@doi [The
  Astrophysical Journal] {10.3847/1538-4357/aaa243}, 853, 105

\bibitem[\protect\citeauthoryear{{Cargill}, {Chen}  \& {Garren}}{{Cargill}
  et~al.}{1994}]{Ca94}
{Cargill} P.~J.,  {Chen} J.,   {Garren} D.~A.,  1994, \mn@doi [\apj]
  {10.1086/173863}, \href
  {https://ui.adsabs.harvard.edu/abs/1994ApJ...423..854C} {423, 854}

\bibitem[\protect\citeauthoryear{Chen, Ma  \& Zhang}{Chen et~al.}{2013}]{Ch13}
Chen H.,  Ma S.,   Zhang J.,  2013, \mn@doi [The Astrophysical Journal]
  {10.1088/0004-637x/778/1/70}, 778, 70

\bibitem[\protect\citeauthoryear{Cheng, Zhang, Ding, Guo  \& Su}{Cheng
  et~al.}{2011}]{Ch11}
Cheng X.,  Zhang J.,  Ding M.~D.,  Guo Y.,   Su J.~T.,  2011, \mn@doi [The
  Astrophysical Journal] {10.1088/0004-637x/732/2/87}, 732, 87

\bibitem[\protect\citeauthoryear{{Cheng} et~al.,}{{Cheng} et~al.}{2014}]{Ch14}
{Cheng} X.,  et~al., 2014, \mn@doi [\apj] {10.1088/0004-637X/780/1/28}, \href
  {http://adsabs.harvard.edu/abs/2014ApJ...780...28C} {780, 28}

\bibitem[\protect\citeauthoryear{{Dasso}, {Nakwacki}, {D{\'e}moulin}  \& {Mand
  rini}}{{Dasso} et~al.}{2007}]{Da07}
{Dasso} S.,  {Nakwacki} M.~S.,  {D{\'e}moulin} P.,   {Mand rini} C.~H.,  2007,
  \mn@doi [\solphys] {10.1007/s11207-007-9034-2}, \href
  {https://ui.adsabs.harvard.edu/abs/2007SoPh..244..115D} {244, 115}

\bibitem[\protect\citeauthoryear{{D{\'e}moulin} \& {Aulanier}}{{D{\'e}moulin}
  \& {Aulanier}}{2010}]{De10}
{D{\'e}moulin} P.,  {Aulanier} G.,  2010, \mn@doi [\apj]
  {10.1088/0004-637X/718/2/1388}, \href
  {https://ui.adsabs.harvard.edu/abs/2010ApJ...718.1388D} {718, 1388}

\bibitem[\protect\citeauthoryear{{Filippov}}{{Filippov}}{2013}]{Fi13}
{Filippov} B.,  2013, \mn@doi [\apj] {10.1088/0004-637X/773/1/10}, \href
  {http://adsabs.harvard.edu/abs/2013ApJ...773...10F} {773, 10}

\bibitem[\protect\citeauthoryear{{Filippov}}{{Filippov}}{2018}]{Fi18}
{Filippov} B.,  2018, \mn@doi [\mnras] {10.1093/mnras/stx3277}, \href
  {https://ui.adsabs.harvard.edu/abs/2018MNRAS.475.1646F} {475, 1646}

\bibitem[\protect\citeauthoryear{{Filippov}}{{Filippov}}{2020}]{Fi20}
{Filippov} B.~P.,  2020, Astronomy Reports, 64, in press

\bibitem[\protect\citeauthoryear{{Filippov} \& {Den}}{{Filippov} \&
  {Den}}{2000}]{Fi00}
{Filippov} B.~P.,  {Den} O.~G.,  2000, \mn@doi [Astronomy Letters]
  {10.1134/1.20397}, \href {http://adsabs.harvard.edu/abs/2000AstL...26..322F}
  {26, 322}

\bibitem[\protect\citeauthoryear{{Filippov} \& {Den}}{{Filippov} \&
  {Den}}{2001}]{Fi01}
{Filippov} B.~P.,  {Den} O.~G.,  2001, \mn@doi [\jgr] {10.1029/2000JA004002},
  \href {http://adsabs.harvard.edu/abs/2001JGR...10625177F} {106, 25177}

\bibitem[\protect\citeauthoryear{{Filippov} \& {Zagnetko}}{{Filippov} \&
  {Zagnetko}}{2008}]{Fi08}
{Filippov} B.,  {Zagnetko} A.,  2008, \mn@doi [Journal of Atmospheric and
  Solar-Terrestrial Physics] {10.1016/j.jastp.2007.08.035}, \href
  {http://adsabs.harvard.edu/abs/2008JASTP..70..614F} {70, 614}

\bibitem[\protect\citeauthoryear{{Forbes} \& {Isenberg}}{{Forbes} \&
  {Isenberg}}{1991}]{Fo91}
{Forbes} T.~G.,  {Isenberg} P.~A.,  1991, \mn@doi [\apj] {10.1086/170051},
  \href {http://adsabs.harvard.edu/abs/1991ApJ...373..294F} {373, 294}

\bibitem[\protect\citeauthoryear{{Forbes} \& {Priest}}{{Forbes} \&
  {Priest}}{1995}]{Fo95}
{Forbes} T.~G.,  {Priest} E.~R.,  1995, \mn@doi [\apj] {10.1086/175797}, \href
  {https://ui.adsabs.harvard.edu/abs/1995ApJ...446..377F} {446, 377}

\bibitem[\protect\citeauthoryear{{Garren} \& {Chen}}{{Garren} \&
  {Chen}}{1994}]{Ga94}
{Garren} D.~A.,  {Chen} J.,  1994, \mn@doi [Physics of Plasmas]
  {10.1063/1.870491}, \href
  {https://ui.adsabs.harvard.edu/abs/1994PhPl....1.3425G} {1, 3425}

\bibitem[\protect\citeauthoryear{{Gary} \& {Moore}}{{Gary} \&
  {Moore}}{2004}]{Ga04}
{Gary} G.~A.,  {Moore} R.~L.,  2004, \mn@doi [\apj] {10.1086/422132}, \href
  {http://adsabs.harvard.edu/abs/2004ApJ...611..545G} {611, 545}

\bibitem[\protect\citeauthoryear{{Gibson} \& {Fan}}{{Gibson} \&
  {Fan}}{2006}]{Gi06}
{Gibson} S.~E.,  {Fan} Y.,  2006, \mn@doi [\apjl] {10.1086/500452}, \href
  {https://ui.adsabs.harvard.edu/abs/2006ApJ...637L..65G} {637, L65}

\bibitem[\protect\citeauthoryear{{Gilbert}, {Holzer}, {Burkepile}  \&
  {Hundhausen}}{{Gilbert} et~al.}{2000}]{Gi00}
{Gilbert} H.~R.,  {Holzer} T.~E.,  {Burkepile} J.~T.,   {Hundhausen} A.~J.,
  2000, \mn@doi [\apj] {10.1086/309030}, \href
  {https://ui.adsabs.harvard.edu/abs/2000ApJ...537..503G} {537, 503}

\bibitem[\protect\citeauthoryear{{Guo}, {Ding}, {Schmieder}, {Li},
  {T{\"o}r{\"o}}  \& {Wiegelmann}}{{Guo} et~al.}{2011}]{Gu11}
{Guo} Y.,  {Ding} M.~D.,  {Schmieder} B.,  {Li} H.,  {T{\"o}r{\"o}} T.,
  {Wiegelmann} T.,  2011, in Astronomical Society of India Conference Series.
  pp 307--313

\bibitem[\protect\citeauthoryear{Holman \& Foord}{Holman \& Foord}{2015}]{Ho15}
Holman G.~D.,  Foord A.,  2015, \mn@doi [The Astrophysical Journal]
  {10.1088/0004-637x/804/2/108}, 804, 108

\bibitem[\protect\citeauthoryear{{Howard} et~al.,}{{Howard}
  et~al.}{2008}]{Ho08}
{Howard} R.~A.,  et~al., 2008, \mn@doi [\ssr] {10.1007/s11214-008-9341-4},
  \href {http://adsabs.harvard.edu/abs/2008SSRv..136...67H} {136, 67}

\bibitem[\protect\citeauthoryear{{Inoue}, {Kusano}, {B{\"u}chner}  \&
  {Sk{\'a}la}}{{Inoue} et~al.}{2018}]{In18}
{Inoue} S.,  {Kusano} K.,  {B{\"u}chner} J.,   {Sk{\'a}la} J.,  2018, \mn@doi
  [Nature Communications] {10.1038/s41467-017-02616-8}, \href
  {https://ui.adsabs.harvard.edu/abs/2018NatCo...9..174I} {9, 174}

\bibitem[\protect\citeauthoryear{{Isenberg} \& {Forbes}}{{Isenberg} \&
  {Forbes}}{2007}]{Is07}
{Isenberg} P.~A.,  {Forbes} T.~G.,  2007, \mn@doi [\apj] {10.1086/522025},
  \href {https://ui.adsabs.harvard.edu/abs/2007ApJ...670.1453I} {670, 1453}

\bibitem[\protect\citeauthoryear{{Ji}, {Wang}, {Schmahl}, {Moon}  \&
  {Jiang}}{{Ji} et~al.}{2003}]{Ji03}
{Ji} H.,  {Wang} H.,  {Schmahl} E.~J.,  {Moon} Y.~J.,   {Jiang} Y.,  2003,
  \mn@doi [\apjl] {10.1086/378178}, \href
  {https://ui.adsabs.harvard.edu/abs/2003ApJ...595L.135J} {595, L135}

\bibitem[\protect\citeauthoryear{Joshi \& Srivastava}{Joshi \&
  Srivastava}{2011}]{Jo11}
Joshi A.~D.,  Srivastava N.,  2011, \mn@doi [The Astrophysical Journal]
  {10.1088/0004-637x/730/2/104}, 730, 104

\bibitem[\protect\citeauthoryear{{Joshi}, {Srivastava}, {Filippov}, {Kayshap},
  {Uddin}, {Chandra}, {Prasad Choudhary}  \& {Dwivedi}}{{Joshi}
  et~al.}{2014}]{Jo14}
{Joshi} N.~C.,  {Srivastava} A.~K.,  {Filippov} B.,  {Kayshap} P.,  {Uddin} W.,
   {Chandra} R.,  {Prasad Choudhary} D.,   {Dwivedi} B.~N.,  2014, \mn@doi
  [\apj] {10.1088/0004-637X/787/1/11}, \href
  {http://adsabs.harvard.edu/abs/2014ApJ...787...11J} {787, 11}

\bibitem[\protect\citeauthoryear{{Kadomtsev}}{{Kadomtsev}}{1966}]{Ka66}
{Kadomtsev} B.~B.,  1966, Reviews of Plasma Physics, \href
  {https://ui.adsabs.harvard.edu/abs/1966RvPP....2..153K} {2, 153}

\bibitem[\protect\citeauthoryear{{Kliem} \& {T{\"o}r{\"o}k}}{{Kliem} \&
  {T{\"o}r{\"o}k}}{2006}]{Kl06}
{Kliem} B.,  {T{\"o}r{\"o}k} T.,  2006, \mn@doi [Physical Review Letters]
  {10.1103/PhysRevLett.96.255002}, \href
  {http://adsabs.harvard.edu/abs/2006PhRvL..96y5002K} {96, 255002}

\bibitem[\protect\citeauthoryear{{Kliem}, {T{\"o}r{\"o}k}, {Titov}, {Lionello},
  {Linker}, {Liu}, {Liu}  \& {Wang}}{{Kliem} et~al.}{2014}]{Kl14}
{Kliem} B.,  {T{\"o}r{\"o}k} T.,  {Titov} V.~S.,  {Lionello} R.,  {Linker}
  J.~A.,  {Liu} R.,  {Liu} C.,   {Wang} H.,  2014, \mn@doi [\apj]
  {10.1088/0004-637X/792/2/107}, \href
  {http://adsabs.harvard.edu/abs/2014ApJ...792..107K} {792, 107}

\bibitem[\protect\citeauthoryear{{Kuperus} \& {Raadu}}{{Kuperus} \&
  {Raadu}}{1974}]{Ku74}
{Kuperus} M.,  {Raadu} M.~A.,  1974, \aap, \href
  {http://adsabs.harvard.edu/abs/1974A/A....31..189K} {31, 189}

\bibitem[\protect\citeauthoryear{{Kuridze}, {Mathioudakis}, {Kowalski}, {Keys},
  {Jess}, {Balasubramaniam}  \& {Keenan}}{{Kuridze} et~al.}{2013}]{Ku13}
{Kuridze} D.,  {Mathioudakis} M.,  {Kowalski} A.~F.,  {Keys} P.~H.,  {Jess}
  D.~B.,  {Balasubramaniam} K.~S.,   {Keenan} F.~P.,  2013, \mn@doi [\aap]
  {10.1051/0004-6361/201220055}, \href
  {http://adsabs.harvard.edu/abs/2013A/A...552A..55K} {552, A55}

\bibitem[\protect\citeauthoryear{{Kushwaha}, {Joshi}, {Veronig}  \&
  {Moon}}{{Kushwaha} et~al.}{2015}]{Ku15}
{Kushwaha} U.,  {Joshi} B.,  {Veronig} A.~M.,   {Moon} Y.-J.,  2015, \mn@doi
  [\apj] {10.1088/0004-637X/807/1/101}, \href
  {http://adsabs.harvard.edu/abs/2015ApJ...807..101K} {807, 101}

\bibitem[\protect\citeauthoryear{{Lemen} et~al.,}{{Lemen} et~al.}{2012}]{Le12}
{Lemen} J.~R.,  et~al., 2012, \mn@doi [\solphys] {10.1007/s11207-011-9776-8},
  \href {http://adsabs.harvard.edu/abs/2012SoPh..275...17L} {275, 17}

\bibitem[\protect\citeauthoryear{{Lepping}, {Jones}  \& {Burlaga}}{{Lepping}
  et~al.}{1990}]{Le90}
{Lepping} R.~P.,  {Jones} J.~A.,   {Burlaga} L.~F.,  1990, \mn@doi [\jgr]
  {10.1029/JA095iA08p11957}, \href
  {https://ui.adsabs.harvard.edu/abs/1990JGR....9511957L} {95, 11957}

\bibitem[\protect\citeauthoryear{Li, Liu, Liu, Elmhamdi  \& Kordi}{Li
  et~al.}{2018}]{Li18}
Li H.,  Liu Y.,  Liu J.,  Elmhamdi A.,   Kordi A.-S.,  2018, \mn@doi
  [Publications of the Astronomical Society of the Pacific]
  {10.1088/1538-3873/aae6a7}, 130, 124401

\bibitem[\protect\citeauthoryear{{Lin}, {Forbes}, {Isenberg}  \&
  {D{\'e}moulin}}{{Lin} et~al.}{1998}]{Li98}
{Lin} J.,  {Forbes} T.~G.,  {Isenberg} P.~A.,   {D{\'e}moulin} P.,  1998,
  \mn@doi [\apj] {10.1086/306108}, \href
  {https://ui.adsabs.harvard.edu/abs/1998ApJ...504.1006L} {504, 1006}

\bibitem[\protect\citeauthoryear{Liu}{Liu}{2008}]{Liu08}
Liu Y.,  2008, \mn@doi [The Astrophysical Journal] {10.1086/589282}, 679, L151

\bibitem[\protect\citeauthoryear{Liu, Su, Xu, Lin, Shibata  \& Kurokawa}{Liu
  et~al.}{2009}]{Liu09}
Liu Y.,  Su J.,  Xu Z.,  Lin H.,  Shibata K.,   Kurokawa H.,  2009, \mn@doi
  [The Astrophysical Journal] {10.1088/0004-637x/696/1/l70}, 696, L70

\bibitem[\protect\citeauthoryear{Liu, Wang, Zhou, Dissauer, Temmer  \& Cui}{Liu
  et~al.}{2018}]{Liu18}
Liu L.,  Wang Y.,  Zhou Z.,  Dissauer K.,  Temmer M.,   Cui J.,  2018, \mn@doi
  [The Astrophysical Journal] {10.3847/1538-4357/aabba2}, 858, 121

\bibitem[\protect\citeauthoryear{{Longcope} \& {Forbes}}{{Longcope} \&
  {Forbes}}{2014}]{Lo14}
{Longcope} D.~W.,  {Forbes} T.~G.,  2014, \mn@doi [\solphys]
  {10.1007/s11207-013-0464-8}, \href
  {https://ui.adsabs.harvard.edu/abs/2014SoPh..289.2091L} {289, 2091}

\bibitem[\protect\citeauthoryear{{Lundquist}}{{Lundquist}}{1951}]{Lu51}
{Lundquist} S.,  1951, \mn@doi [Physical Review] {10.1103/PhysRev.83.307},
  \href {https://ui.adsabs.harvard.edu/abs/1951PhRv...83..307L} {83, 307}

\bibitem[\protect\citeauthoryear{{McCauley}, {Su}, {Schanche}, {Evans}, {Su},
  {McKillop}  \& {Reeves}}{{McCauley} et~al.}{2015}]{Mc15}
{McCauley} P.~I.,  {Su} Y.~N.,  {Schanche} N.,  {Evans} K.~E.,  {Su} C.,
  {McKillop} S.,   {Reeves} K.~K.,  2015, \mn@doi [\solphys]
  {10.1007/s11207-015-0699-7}, \href
  {https://ui.adsabs.harvard.edu/abs/2015SoPh..290.1703M} {290, 1703}

\bibitem[\protect\citeauthoryear{{Molodenskii} \& {Filippov}}{{Molodenskii} \&
  {Filippov}}{1987}]{Mo87}
{Molodenskii} M.~M.,  {Filippov} B.~P.,  1987, \sovast, \href
  {http://adsabs.harvard.edu/abs/1987SvA....31..564M} {31, 564}

\bibitem[\protect\citeauthoryear{{Myers}, {Yamada}, {Ji}, {Yoo}, {Fox},
  {Jara-Almonte}, {Savcheva}  \& {Deluca}}{{Myers} et~al.}{2015}]{My15}
{Myers} C.~E.,  {Yamada} M.,  {Ji} H.,  {Yoo} J.,  {Fox} W.,  {Jara-Almonte}
  J.,  {Savcheva} A.,   {Deluca} E.~E.,  2015, \mn@doi [\nat]
  {10.1038/nature16188}, \href
  {https://ui.adsabs.harvard.edu/abs/2015Natur.528..526M} {528, 526}

\bibitem[\protect\citeauthoryear{{Myers}, {Yamada}, {Ji}, {Yoo}, {Jara-Almonte}
   \& {Fox}}{{Myers} et~al.}{2016}]{My16}
{Myers} C.~E.,  {Yamada} M.,  {Ji} H.,  {Yoo} J.,  {Jara-Almonte} J.,   {Fox}
  W.,  2016, \mn@doi [Physics of Plasmas] {10.1063/1.4966691}, \href
  {https://ui.adsabs.harvard.edu/abs/2016PhPl...23k2102M} {23, 112102}

\bibitem[\protect\citeauthoryear{{Olmedo} \& {Zhang}}{{Olmedo} \&
  {Zhang}}{2010}]{Ol10}
{Olmedo} O.,  {Zhang} J.,  2010, \mn@doi [\apj] {10.1088/0004-637X/718/1/433},
  \href {https://ui.adsabs.harvard.edu/abs/2010ApJ...718..433O} {718, 433}

\bibitem[\protect\citeauthoryear{{Patsourakos}, {Vourlidas}  \&
  {Stenborg}}{{Patsourakos} et~al.}{2013}]{Pa13}
{Patsourakos} S.,  {Vourlidas} A.,   {Stenborg} G.,  2013, \mn@doi [\apj]
  {10.1088/0004-637X/764/2/125}, \href
  {http://adsabs.harvard.edu/abs/2013ApJ...764..125P} {764, 125}

\bibitem[\protect\citeauthoryear{{Pesnell}, {Thompson}  \&
  {Chamberlin}}{{Pesnell} et~al.}{2012}]{Pe12}
{Pesnell} W.~D.,  {Thompson} B.~J.,   {Chamberlin} P.~C.,  2012, \mn@doi
  [\solphys] {10.1007/s11207-011-9841-3}, \href
  {http://adsabs.harvard.edu/abs/2012SoPh..275....3P} {275, 3}

\bibitem[\protect\citeauthoryear{{Priest} \& {Forbes}}{{Priest} \&
  {Forbes}}{1990}]{Pr90}
{Priest} E.~R.,  {Forbes} T.~G.,  1990, \mn@doi [\solphys]
  {10.1007/BF00153054}, \href
  {http://adsabs.harvard.edu/abs/1990SoPh..126..319P} {126, 319}

\bibitem[\protect\citeauthoryear{{Sarkar} \& {Srivastava}}{{Sarkar} \&
  {Srivastava}}{2018}]{Sa18}
{Sarkar} R.,  {Srivastava} N.,  2018, \mn@doi [\solphys]
  {10.1007/s11207-017-1235-8}, \href
  {https://ui.adsabs.harvard.edu/abs/2018SoPh..293...16S} {293, 16}

\bibitem[\protect\citeauthoryear{{Schou} et~al.,}{{Schou} et~al.}{2012}]{Sc12}
{Schou} J.,  et~al., 2012, \mn@doi [\solphys] {10.1007/s11207-011-9842-2},
  \href {http://adsabs.harvard.edu/abs/2012SoPh..275..229S} {275, 229}

\bibitem[\protect\citeauthoryear{{Shafranov}}{{Shafranov}}{1966}]{Sh66}
{Shafranov} V.~D.,  1966, Reviews of Plasma Physics, \href
  {http://adsabs.harvard.edu/abs/1966RvPP....2..103S} {2, 103}

\bibitem[\protect\citeauthoryear{{Sun}}{{Sun}}{2013}]{Su13}
{Sun} X.,  2013, arXiv e-prints, \href
  {https://ui.adsabs.harvard.edu/abs/2013arXiv1309.2392S} {p. arXiv:1309.2392}

\bibitem[\protect\citeauthoryear{Sun et~al.,}{Sun et~al.}{2015}]{Su15}
Sun X.,  et~al., 2015, \mn@doi [The Astrophysical Journal]
  {10.1088/2041-8205/804/2/l28}, 804, L28

\bibitem[\protect\citeauthoryear{{T{\"o}r{\"o}k} \& {Kliem}}{{T{\"o}r{\"o}k} \&
  {Kliem}}{2005}]{To05}
{T{\"o}r{\"o}k} T.,  {Kliem} B.,  2005, \mn@doi [\apjl] {10.1086/462412}, \href
  {http://adsabs.harvard.edu/abs/2005ApJ...630L..97T} {630, L97}

\bibitem[\protect\citeauthoryear{{T{\"o}r{\"o}k}, {Kliem}  \&
  {Titov}}{{T{\"o}r{\"o}k} et~al.}{2004}]{To04}
{T{\"o}r{\"o}k} T.,  {Kliem} B.,   {Titov} V.~S.,  2004, \mn@doi [\aap]
  {10.1051/0004-6361:20031691}, \href
  {https://ui.adsabs.harvard.edu/abs/2004A&A...413L..27T} {413, L27}

\bibitem[\protect\citeauthoryear{{Tsap}, {Filippov}  \& {Kopylova}}{{Tsap}
  et~al.}{2019}]{Ts19}
{Tsap} Y.~T.,  {Filippov} B.~P.,   {Kopylova} Y.~G.,  2019, \mn@doi [\solphys]
  {10.1007/s11207-019-1423-9}, \href
  {https://ui.adsabs.harvard.edu/abs/2019SoPh..294...35T} {294, 35}

\bibitem[\protect\citeauthoryear{Vasantharaju, Vemareddy, Ravindra  \&
  Doddamani}{Vasantharaju et~al.}{2018}]{Va18}
Vasantharaju N.,  Vemareddy P.,  Ravindra B.,   Doddamani V.~H.,  2018, \mn@doi
  [The Astrophysical Journal] {10.3847/1538-4357/aac272}, 860, 58

\bibitem[\protect\citeauthoryear{Wang, Liu, Wang, Liu, Chen, Liu, Zhou  \&
  Zhang}{Wang et~al.}{2017}]{Wa17}
Wang D.,  Liu R.,  Wang Y.,  Liu K.,  Chen J.,  Liu J.,  Zhou Z.,   Zhang M.,
  2017, \mn@doi [The Astrophysical Journal] {10.3847/2041-8213/aa79f0}, 843, L9

\bibitem[\protect\citeauthoryear{Wang, Liu, Wang, Gou, Zhang, Zhou  \&
  Zhang}{Wang et~al.}{2018}]{Wa18}
Wang D.,  Liu R.,  Wang Y.,  Gou T.,  Zhang Q.,  Zhou Z.,   Zhang M.,  2018,
  \mn@doi [The Astrophysical Journal] {10.3847/1538-4357/aaef35}, 869, 177

\bibitem[\protect\citeauthoryear{{Wuelser} et~al.,}{{Wuelser}
  et~al.}{2004}]{Wu04}
{Wuelser} J.-P.,  et~al., 2004, in {Fineschi} S.,  {Gummin} M.~A.,  eds,
  Society of Photo-Optical Instrumentation Engineers (SPIE) Conference Series
  Vol. 5171, Society of Photo-Optical Instrumentation Engineers (SPIE)
  Conference Series. pp 111--122, \mn@doi{10.1117/12.506877}

\bibitem[\protect\citeauthoryear{Xue, Yan, Zhao, Xiang, Yang  \& Guo}{Xue
  et~al.}{2016}]{Xu16}
Xue Z.,  Yan X.,  Zhao L.,  Xiang Y.,  Yang L.,   Guo Y.,  2016, \mn@doi
  [Publications of the Astronomical Society of Japan] {10.1093/pasj/psv113}, 68

\bibitem[\protect\citeauthoryear{{Zaitsev} \& {Stepanov}}{{Zaitsev} \&
  {Stepanov}}{2018}]{Za18}
{Zaitsev} V.~V.,  {Stepanov} A.~V.,  2018, \mn@doi [Journal of Atmospheric and
  Solar-Terrestrial Physics] {10.1016/j.jastp.2018.06.004}, \href
  {https://ui.adsabs.harvard.edu/abs/2018JASTP.179..149Z} {179, 149}

\bibitem[\protect\citeauthoryear{Zhou, Cheng, Zhang, Wang, Wang, Liu, Zhuang
  \& Cui}{Zhou et~al.}{2019}]{Zh19}
Zhou Z.,  Cheng X.,  Zhang J.,  Wang Y.,  Wang D.,  Liu L.,  Zhuang B.,   Cui
  J.,  2019, \mn@doi [The Astrophysical Journal] {10.3847/2041-8213/ab21cb},
  877, L28

\bibitem[\protect\citeauthoryear{{van Tend} \& {Kuperus}}{{van Tend} \&
  {Kuperus}}{1978}]{va78}
{van Tend} W.,  {Kuperus} M.,  1978, \mn@doi [\solphys] {10.1007/BF00154935},
  \href {http://adsabs.harvard.edu/abs/1978SoPh...59..115V} {59, 115}

\makeatother
\end{thebibliography}

%

\bsp	
\label{lastpage}
\end{document}